\documentclass[aps,prl,reprint,letterpaper,amsmath,amssymb,superscriptaddress,twocolumn,preprintnumbers,longbibliography]{revtex4-1}
\usepackage{graphicx}
\usepackage{amsmath}
\usepackage[dvipsnames]{xcolor}
\newcommand{\delV}[1]{\textcolor{red}{}}

\begin{document}
\title{Electron and hole spin relaxation in InP-based self-assembled quantum dots emitting
at telecom wavelengths}
\author{A.~V.~Mikhailov}
\affiliation{Experimentelle Physik 2, Technische Universit\"{a}t Dortmund, D-44221 Dortmund, Germany}
\affiliation{Spin Optics Laboratory, St. Petersburg State University, 199034 St. Petersburg, Russia}
\author{V.~V.~Belykh}
\email[]{vasilii.belykh@tu-dortmund.de}
\affiliation{Experimentelle Physik 2, Technische Universit\"{a}t Dortmund, D-44221 Dortmund, Germany}
\affiliation{P.N. Lebedev Physical Institute of the Russian Academy of Sciences, 119991 Moscow, Russia}
\author{D.~R.~Yakovlev}
\affiliation{Experimentelle Physik 2, Technische Universit\"{a}t Dortmund, D-44221 Dortmund, Germany}
\affiliation{Ioffe Institute, Russian Academy of Sciences, 194021 St. Petersburg, Russia}
\author{P.~S.~Grigoryev}
\affiliation{Experimentelle Physik 2, Technische Universit\"{a}t Dortmund, D-44221 Dortmund, Germany}
\affiliation{Spin Optics Laboratory, St. Petersburg State University, 199034 St. Petersburg, Russia}
\author{J.~P.~Reithmaier}
\affiliation{Institute of Nanostructure Technologies and Analytics (INA), CINSaT, University of Kassel, D-34132 Kassel, Germany}
\author{M.~Benyoucef}
\affiliation{Institute of Nanostructure Technologies and Analytics (INA), CINSaT, University of Kassel, D-34132 Kassel, Germany}
\author{M.~Bayer}
\affiliation{Experimentelle Physik 2, Technische Universit\"{a}t Dortmund, D-44221 Dortmund, Germany}
\affiliation{Ioffe Institute, Russian Academy of Sciences, 194021 St. Petersburg, Russia}

\begin{abstract}
We investigate the electron and hole spin relaxation in an ensemble of self-assembled InAs/In$_{0.53}$Al$_{0.24}$Ga$_{0.23}$As/InP quantum dots with emission wavelengths around $1.5$~$\mu$m by pump-probe Faraday rotation spectroscopy. Electron spin dephasing due to the randomly oriented nuclear Overhauser fields is observed. At low temperatures we find a sub-microsecond longitudinal electron spin relaxation time $T_1$ which unexpectedly strongly depends on temperature. At high temperatures the electron spin relaxation time is limited by optical phonon scattering through spin-orbit interaction decreasing down to $0.1$~ns at 260~K. We show that the hole spin relaxation is activated much more effectively by a temperature increase compared to the electrons.
\end{abstract}
\maketitle

\section{Introduction}
Semiconductor quantum dots (QDs) offer a  promising platform for quantum information technologies \cite{Loss1998}. An electron spin in a QD, often considered as a candidate for a quantum bit (Qubit), can be efficiently manipulated by light pulses which gives the possibility of easy integration into existing optical telecommunication networks. In this respect QDs emitting in the telecom spectral range (1.3-1.6 $\mu$m) are especially attractive
\cite{Fafard1996,Ponchet1995,Taskinen1997,Paranthoen2001,Saito2001,Wang2001,Lelarge2005,VanVeldhoven2009,Liao2017,Sapienza2016,Mrowinski2016}. In particular, potential applications as laser active medium \cite{Saito2001,Wang2001,Lelarge2005}, single-photon emitters \cite{Miyazawa2005,Takemoto2007,Takemoto2010,Birowosuto2012,Benyoucef2013,Liu2013,Dusanowski2014,Kim2016,Paul2017}, and polarization-entangled photons emitters \cite{Olbrich2017} are envisaged.

While for III-V QDs emitting in the technologically and spectroscopically easily accessible wavelength range below 1 $\mu$m the spin properties have been intensively studied in recent decades \cite{Dyakonov2017,Bayer1999,Kroutvar2004,Dutt2005,Greilich2006,Hernandez2008,Fras2012,Dou2012,Semenov2007,Braun2005,Eble2009,Bechtold2015}, information on the spin dynamics of QDs emitting at longer wavelengths, in particular in the telecom spectral range, is limited. So far, the electron and hole $g$ factors \cite{Belykh2015} with record-high anisotropies \cite{Belykh2016,VanBree2016,Belykh2016a} were measured. The dynamics of the photoluminescence polarization degree related to the exciton spin dynamics was measured as well \cite{Syperek2016}.

In this paper we address the spin lifetimes of carriers in InAs/In$_{0.53}$Al$_{0.24}$Ga$_{0.23}$As/InP QDs emitting at telecom wavelengths which have not been measured so far to the best of our knowledge. At weak magnetic fields, the spin dynamics of the resident electrons in the QDs is governed by the hyperfine interaction with the nuclei and the regime described theoretically in Ref.~\cite{Merkulov2002} is observed. In increased longitudinal magnetic fields, at low temperatures we observe a sub-microsecond decay of spin polarization. With increasing temperature we observe a drastic decrease of both $T_1$ and $T_2^*$ which at high temperatures is mediated by the electron interaction with LO phonons.

\section{Experimental details}
The QD sample was grown by molecular-beam epitaxy on an (100)-oriented InP substrate. The nominally undoped QDs were formed from 5.5 monolayers of InAs sandwiched between In$_{0.53}$Al$_{0.24}$Ga$_{0.23}$As barriers. The optically active QDs have a diameter of $\sim 50$~nm and a height of $\sim 10$~nm; their density is about $10^{10}$ cm$^{-2}$.

The sample is held at temperatures in the range 5-260~K in a helium bath cryostat with a split-coil superconducting magnet. Magnetic fields up to $B=2$~T are applied either in Faraday (parallel to the light propagation direction, coinciding with the sample growth axis) or in Voigt (perpendicular to the light propagation direction) geometry. Pump-probe Faraday rotation is employed to study the carriers spin dynamics in the QDs \cite{Dyakonov2017}. Two laser systems are used. The first one consists of a mode-locked Yb:KGW laser pumping an optical parametric amplifier and has the pulse repetition frequency of $40$~MHz (repetition period $T_\text{R}=25$~ns). The second laser system is composed of a pulsed Ti:Saphire laser pumping an optical parametric oscillator (OPO) and has the pulse repetition rate of 76~MHz ($T_\text{R}=13$~ns). The spectral width of the laser systems was shaped below 20~nm, centered at 1520 nm which matches the QDs luminescence spectrum \cite{Belykh2015,Belykh2016}. The pulse duration for both systems is less than $2$~ps.

The laser beam is split into the pump and the probe. The pump pulses are circularly polarized and create carrier spin polarization in the QDs. The mechanism of optical spin orientation in QDs is described in Refs.\cite{Dyakonov2017,Smirnov2018}. The carriers' spin polarization is analysed by measuring the Faraday ellipticity of the initially linearly polarized probe pulses after transmission through the sample \cite{Yugova2009}. Varying the time delay between the pump and probe pulses by a mechanical delay line gives the time dependence of the spin polarization. The polarization of the pump beam is modulated between $\sigma^+$ and $\sigma^-$ by a photo-elastic modulator for synchronous detection.



\section{Results and discussion}
\begin{figure}
\includegraphics[width=1\linewidth]{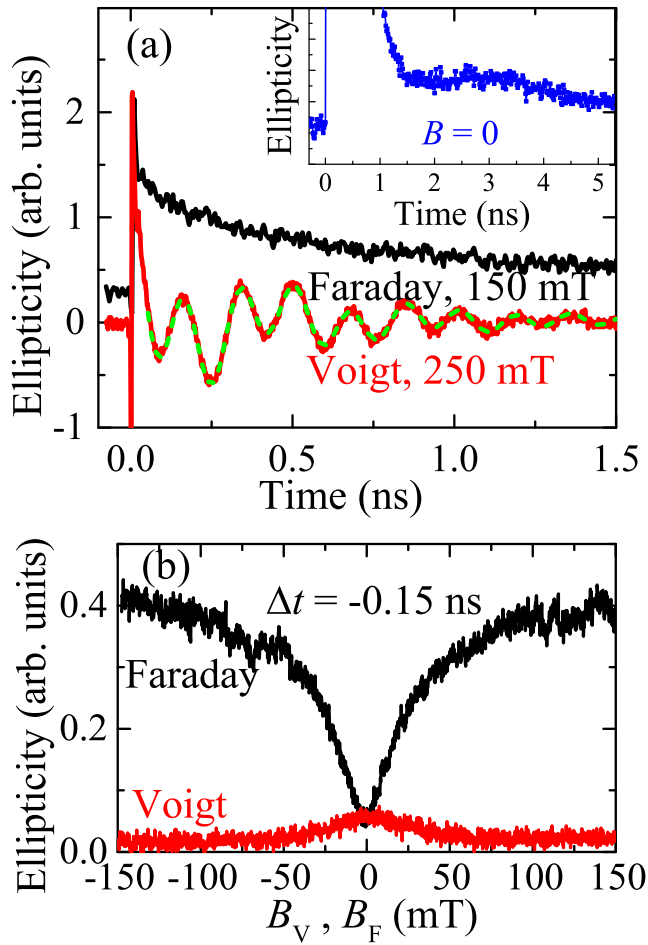}%
\caption{(a) Dynamics of Faraday ellipticity signal for the magnetic field applied in Faraday (black line) and Voigt (red line) geometry, $T_\text{R}=25$~ns. Green dashed line shows fit to the experimental data with two exponentially decaying oscillating functions. Inset shows dynamics at zero magnetic field with the focus on its nonmonotonic behavior related to electron spin precession in the nuclear Overhauser field. (b) Ellipticity signal as function of magnetic field applied in Faraday (black line) and Voigt (red line) geometry at $t = -0.15$~ns, $T_\text{R}=13$~ns. $T = 5-10$~K.}
\label{fig:VoigtFara}
\end{figure}

Figure~\ref{fig:VoigtFara}(a) shows the dynamics of the Faraday ellipticity signal for the magnetic field applied in Voigt ($B_\text{V}$) and Faraday ($B_\text{F}$) geometries. In the Voigt geometry the dynamics is composed of the sum of two decaying oscillatory functions, the corresponding fit is shown by the green dashed line. The two oscillation components correspond to the electron and the hole spin precession with $g$ factors $\lvert g_\text{e}\rvert=1.7$ and $\lvert g_\text{h}\rvert=0.7$, respectively. This attribution is based on the detailed investigation of the anisotropies and energy dependencies of these $g$ factors, as well as on theoretical calculations \cite{Belykh2015,Belykh2016}. Note, that the signal may be contributed by charged and neutral QDs. In the latter case, the exchange interaction between electron and hole will lead to a nonlinear magnetic field dependence of the precession frequency \cite{Yugova2007} as well as zero-field spin beats and fast spin dephasing due to inhomogeneous fine structure splitting \cite{Tartakovskii2004}. Both effects are not observed in the experiment \cite{Belykh2015} indicating either an exchange interaction lower than 1~$\mu$eV or/and that the signal is contributed by charged QDs (the most probable scenario). In both cases we observe uncoupled precessions of the electron and hole spins. The dephasing time $T_2^*$ of each of the carrier spin polarizations at low temperatures is determined by the random nuclear Overhauser fields $\mathbf{B}_\text{N}$ (which act on the carrier spins via the hyperfine interaction) if the external magnetic field $B_\text{V}$ is smaller than $B_\text{N}$. At higher $B_\text{V}$, on the other hand, the time $T_2^*$ is determined by the spread of the $g$-factors $\delta g$ in the QD ensemble \cite{Dyakonov2017}:
\begin{subequations}
\begin{align}
1/T_2^* \approx |g|\mu_\text{B} B_\text{N}/\hbar,~~~~~~B_\text{V} \lesssim B_\text{N}, \label{eq:T2Nucl}\\
1/T_2^* \approx \delta g\mu_\text{B} B_\text{V}/\hbar,~~~~~~B_\text{V} \gg B_\text{N}. \label{eq:T2g}
\end{align}
\end{subequations}
Therefore, as it was shown for the same QDs in Refs.~\cite{Belykh2015,Belykh2016}, an increase in $B_\text{V}$ causes a decrease of the signal decay time.

In a longitudinal magnetic field $B_\text{F}$, the spin polarization decays monotonically without oscillations \cite{Belykh2016,Belykh2016b}. In high enough $B_\text{F}$, this decay is characterized by a fast component with decay time $0.6$~ns, close to the exciton lifetime, and a slow component of somewhat smaller amplitude. The existence of the long-living component indicates the presence of \emph{resident} charge carriers in a fraction of the QD ensemble. The decay of the slow component is determined by the longitudinal spin relaxation time $T_1$, which at low temperatures exceeds the period between the laser pulses $T_\text{R}$. This leads to an accumulation of spin polarization, and a signal offset appears at negative pump-probe delays, which can be identified in the Faraday geometry data in Fig.~\ref{fig:VoigtFara}(a). Note that this offset is absent in Voigt geometry.

The effects of longitudinal and transverse magnetic fields on the carrier spin polarization at a small negative delay $\Delta t = -0.15$~ns (which is equivalent to a large delay $t\approx T_\text{R}$ after the previous laser pulse) are presented in Fig.~\ref{fig:VoigtFara}(b), see the black and red lines, respectively. When the longitudinal magnetic field is scanned, the signal has a minimum at $B_\text{F} = 0$ and develops then into a polarization recovery curve (PRC). At high enough longitudinal magnetic field, the PRC signal is mainly determined by the longitudinal spin relaxation time $T_1$.

At zero field the decay of the total spin polarization is governed by the nuclear fields. For an arbitrary QD the direction of the total nuclear Overhauser field $\textbf{B}_\text{N}$ acting on an electron (a hole) spin is random. The electron spin component perpendicular to $\textbf{B}_\text{N}$ precesses around $\textbf{B}_\text{N}$. When precessing, this spin component is averaged over all QDs, and it decays on a short timescale given by Eq.~\eqref{eq:T2Nucl}. On the other hand, the electron spin component along $\textbf{B}_\text{N}$ decays during the much longer time $T_1$. Averaging over all QDs, and, thus, over all directions of $\textbf{B}_\text{N}$, shows that the nonprecessing component amounts to $1/3$ of the initial electron spin polarization \cite{Merkulov2002,Braun2005,Eble2009}. When the \emph{longitudinal} magnetic field $B_\text{F}$ is increased, the nonprecessing spin component along $\textbf{B}_\text{F} + \textbf{B}_\text{N}$ is increased, leading to an increase of the Faraday ellipticity signal due to accumulation of the long-living spin polarization. This simplified picture predicts the drop in the PRC curve [Fig.~\ref{fig:VoigtFara}(b)] at $B_\text{F}=0$ to $1/3$ from the signal at high $B_\text{F}$ and the half width at half maximum (HWHM) of the PRC curve to be equal to $B_\text{N}$. However, a more detailed analysis should take into account the random nuclear spin precession due to the quadrupole splitting which is especially large in the studied QDs having large strain. The effect of the nuclear spin evolution on $T_1$ and on the shape of the PRC curve was considered in Ref.~\cite{Smirnov2018}. It was shown that a decreased correlation time $\tau_\text{c}$ of the nuclear spin evolution, which for QDs typically is in the submicrosecond range, leads to (i) shortening of $T_1$, (ii) increase of the PRC dropdown amplitude, (iii) increase of the PRC width. We observe all three effects in the experiment. Indeed, (i) a decreased $\tau_\text{c}$  with respect to that in standard QDs emitting at shorter wavelengths, results in a decreased $T_1$ \cite{Zhukov2018}, (ii) the spin polarization at $B_\text{F}=0$ is smaller than $1/3$ of its value at high longitudinal magnetic fields (it amounts to $\sim 1/8$ only), (iii) the nuclear field estimated from the electron $T_2^*$ at zero external magnetic field using Eq.~\eqref{eq:T2Nucl} (11~mT) is smaller than the 30~mT HWHM of the PRC curve. A similar HWHM of the PRC curve was observed for negatively charged QDs emitting at shorter wavelength, while the corresponding HWHM for positively charged QDs is about 10 times smaller \cite{Zhukov2018}.

When the \emph{transverse} magnetic field $B_\text{V}$ is increased, it contributes to the nuclear field and the nonprecessing spin polarization (directed along $\textbf{B}_\text{V} + \textbf{B}_\text{N}$) has a vanishing projection on the probe beam direction. This leads to the decrease of the ellipticity signal at negative pump-probe delays with increasing $B_\text{V}$ [Fig.~\ref{fig:VoigtFara}(b), red line].
The half-width of the zero-field peak in this curve gives the value $\approx 30$~mT, the same as in PRC. Note that there are no resonant spin amplification \cite{Kikkawa1998} peaks at nonzero $B_\text{V}$ due to the short transverse dephasing time $T_2^*$ [see Eqs.~(\ref{eq:T2Nucl}-\ref{eq:T2g})]. Another feature characteristic for an electron spin subject to nuclear fields is the non-monotonic decay of the spin polarization in zero external magnetic field, showing a local minimum at the time of $(2-3)T_2^*$, where $T_2^*$ can be estimated from Eq.~\eqref{eq:T2Nucl} \cite{Merkulov2002,Bechtold2015}. This feature agrees with our observations at $t \approx 1.8$~ns [see the inset in Fig.~\ref{fig:VoigtFara}(a)], despite the small amplitude of the minimum.

\begin{figure}[h]
\includegraphics[width=\linewidth]{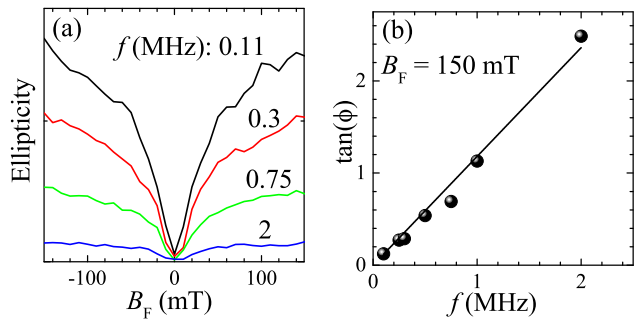}%
\caption{\label{fig:SpinInert} Spin inertia effect. (a) Polarization recovery curves taken at a small negative delay for different pump modulation frequencies. The curves are vertically shifted to the same value at $B_\text{F} = 0$. (b) Frequency dependence of $\tan(\phi)$, where $\phi$ is the retardation phase of the spin polarization modulation with respect to the pump modulation at $B_\text{F}=150$~mT. The line shows a linear fit. $T = 5$~K, $T_\text{R}=13$~ns.}
\end{figure}

Next, we concentrate on evaluating the longitudinal spin relaxation time $T_1$ and studying its temperature dependence.
At low temperatures we use the spin inertia method \cite{Heisterkamp2015} for that purpose. In order to perform synchronous detection, the intensity of the circularly-polarized pump is modulated at frequency $f$: $P = P_0[1+\cos(2\pi f t)]/2$. When the modulation period $1/f$ exceeds the $T_1$ time, the accumulated spin polarization is modulated from 0 up to the maximal value determined by the pumping rate $P_0$ and $T_1$. When the modulation frequency is increased, so that $1/f$ becomes comparable to $T_1$, the spin polarization decreases. One can show that in the case $T_1 \gg T_\text{R}, 1/f \gg T_\text{R}$ the accumulated spin polarization is given by:
\begin{eqnarray}
\label{eq:SISol}
S(t)=P_0 T_1 + A \cos(2\pi f t - \phi),\\
\label{eq:A}
A = \frac{P_0 T_1}{\sqrt{1+(2\pi T_1 f)^2}},\\
\label{eq:phi}
\tan(\phi) = 2 \pi T_1 f.
\end{eqnarray}
Thus, with increasing $f$, the modulation of the spin polarization decreases in amplitude $A$ [Eq.~\eqref{eq:A}] and becomes retarded relative to the pump modulation by the phase $\phi$  [Eq.~\eqref{eq:phi}]. By performing synchronous detection on the pumping frequency $f$, we are able to measure both $A$ and $\phi$. Figure~\ref{fig:SpinInert}(a) shows PRCs measured at different pump modulation frequencies. The ellipticity, reflecting the amplitude of the spin polarization, indeed decreases with increasing $f$. Figure~\ref{fig:SpinInert}(b) shows that $\tan(\phi)$ increases almost linearly with $f$, in agreement with Eq.~\eqref{eq:phi}, allowing us to estimate $T_1\approx 190$~ns from Eq.~\eqref{eq:phi} at $B_\text{F} = 150$~mT.

Let us make two remarks about the spin inertia method and the validity of Eqs.~(\ref{eq:SISol}-\ref{eq:phi}). First, we do not take into account the saturation effect in QDs: at high-enough pump powers and long-enough $T_1$, the majority of QDs becomes spin-polarized and is no longer affected by further pumping. A more detailed analysis shows that the saturation leads to an effective shortening of $T_1$ entering into Eqs.~(\ref{eq:SISol}-\ref{eq:phi}). Second, the above analysis assumes a monoexponential dynamics of the spin polarization, characterized by a single time $T_1$. One can show that in the case of a more complex dynamics the dependence of the spin polarization modulation amplitude on $f$ [Eq.~\eqref{eq:A}] is dominated by the slow component, while the frequency dependence of the retardation phase $\phi$ [Eq.~\eqref{eq:phi}] is dominated by the fast component. Thus, the estimated value $T_1 = 190$~ns is the lower limit for the decay time of the fast component in the longitudinal spin polarization dynamics.

\begin{figure}[h]
\includegraphics[width=1.\linewidth]{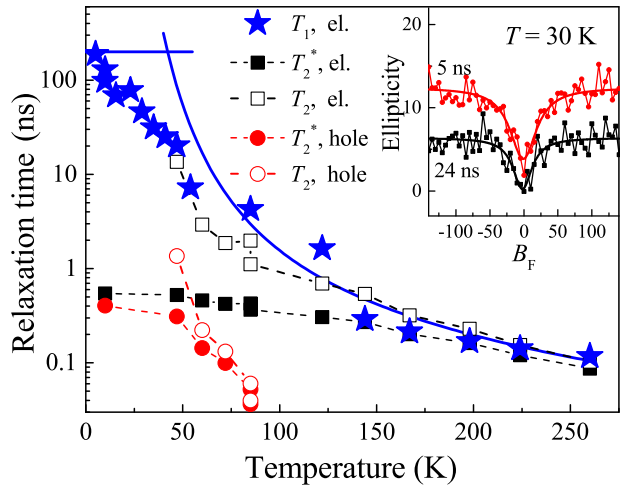}%
\caption{\label{fig:t1t2} Temperature dependencies of longitudinal electron spin relaxation time $T_1$ (stars), inhomogeneous transverse spin relaxation times $T_2^*$ for electrons and holes (solid squares and circles, respectively), and homogeneous transverse spin relaxation times $T_2$ for electrons and holes (open squares and circles, respectively). Solid lines show zero-temperature value of $T_1$ and $T_1$ determined by relaxation with LO phonons according to Eq.~\eqref{eq:LO}. Dashed lines are guides to the eye. Inset shows polarization recovery curves measured at different time delays (symbols) fitted with Lorentzians (solid lines) at $T = 30$~K.}
\end{figure}

The temperature dependence of the longitudinal spin relaxation time $T_1$ is shown in Fig.~\ref{fig:t1t2} by stars. At $T \leq 10$~K, the times $T_1$ are determined by the spin inertia method. For $15 \leq T \leq 50$~K, where $T_1$ becomes comparable to the laser repetition period $T_\text{R}$, they are extracted from the ratio of the ellipticity signals in the PRC curves at different time delays $S(t+\Delta t)/S(t) = \exp(-\Delta t/T_1)$ (inset in Fig.~\ref{fig:t1t2}). Note that this ratio weakly depends on $|B_\text{F}| < 150$~mT, and for final $T_1$ determination we took the ratio of the PRC dip depths. For higher temperatures ($T > 50$~K), where $T_1 < T_\text{R}$, $T_1$ was directly determined as the decay time of the slow component of the ellipticity signal at $B_\text{F} = 150$~mT.

It is instructive to compare the temperature dependence of $T_1$ to that of $T_2^*$ for electrons and holes (Fig.~\ref{fig:t1t2}, solid squares and circles, respectively). The times $T_2^*$ for electrons and holes are determined from the decay of two oscillating components in the transverse magnetic field $B_\text{V} = 250$~mT [see Fig.~\ref{fig:VoigtFara}(a)]. At low temperatures $T_2^*$ is determined by the nuclear field and the spread of $g$ factors [Eqs.~(\ref{eq:T2Nucl},\ref{eq:T2g})] and is much shorter than $T_1$. At higher temperatures the homogeneous dephasing mechanisms related to phonons become important and $T_2^*$ decreases with $T$. One can separate the inhomogeneous ($T_2^\text{inh}$) and homogeneous ($T_2$) contributions to $T_2^*$:
\begin{equation}
1/T_2^*=1/T_2^\text{inh}+1/T_2.
\label{eq:T2}
\end{equation}
Note that $T_2^\text{inh}$ is almost temperature independent as evidenced from the temperature independent width of the PRC minimum (for $T<50$~K where it can be measured) which is determined by the nuclear fields and from the temperature-independent spread of $g$ factors, which is determined by the QDs' shape and composition spread. Taking into account that $T_2^\text{inh} \approx T_2^*(T=0)$, we can estimate the homogeneous transverse spin relaxation times $T_2$ at elevated temperatures using Eq.~\eqref{eq:T2} (at least the part of the homogeneous spin relaxation rate that is temperature dependent). They are shown by the open squares and circles in Fig.~\ref{fig:t1t2} for electrons and holes, respectively. The decrease of $T_2$ with increasing temperature is especially pronounced for holes. For electrons, $T_2$ is close to $T_1$ in the whole temperature range, as it was predicted theoretically \cite{Golovach2004}, which allows us to attribute the $T_1$ dependence and, in general, the long-living spin polarization component to electrons that are resident in a fraction of QDs.

Now we discuss the origin of the $T_1$ temperature dependence for the electrons. In the limit of zero temperature $T_1$ is determined by the hyperfine interaction with nuclear spins as already discussed (horizontal solid line in Fig.~\ref{fig:t1t2}).
For sufficiently high temperatures, $T \gtrsim 50$~K, the spin relaxation is governed by the interaction with LO phonons \cite{Tsitsishvili2002Temp,Dou2012,Flissikowski2003}. In particular, the two-phonon mechanism with absorption and emission of an optical phonon leads to spin relaxation. The relaxation rate due to this process can be described by the following equation \cite{Tsitsishvili2002Temp,Dou2012}:
\begin{multline}
     1/T_\text{1,LO}=\beta N_\text{LO} (N_\text{LO}+1), \\ N_\text{LO}=\bigl[\exp(\epsilon_\text{LO}/k_\text{B}T)-1\bigr]^{-1},
     \label{eq:LO}
\end{multline}
where $N_\text{LO}$ is the number of phonons, $\epsilon_\text{LO}$ is LO phonon energy, and $\beta$ defines the strength of electron-phonon interaction. The corresponding dependence shown in Fig.~\ref{fig:t1t2} by the solid line with $\epsilon_\text{LO} = 30$~meV (LO phonon energy in InAs) and $\beta = 20$~ns$^{-1}$ fits the experimental data at high temperatures.

The relatively strong temperature dependence of $T_1$ for low temperatures, $T \lesssim 50$~K, is unclear. The usual temperature-dependent QD spin relaxation mechanisms, spin-orbit interaction involving phonons \cite{Khaetskii2001} and phonon-activated electron-nuclear flip-flop processes \cite{Khaetskii2001,Erlingsson2001,Abalmassov2004}, give rates several orders smaller than in the experiment. We note that for the QDs emitting around 900~nm a similar temperature dependence of $T_2$ was reported \cite{Hernandez2008}. However, in that case the $T_2$ variation starts from $T = 15$~K, while in our case $T_1$ strongly depends on $T$ already from 5~K. One possible source of temperature-dependent spin relaxation at low temperatures might be exchange interaction with carriers in the wetting layer which are localized by shallow inhomogeneities. With a moderate increase of temperature these carriers become delocalized activating exchange interaction.

\section{Conclusion}
To summarize, we have studied the longitudinal and transverse electron spin relaxation in an ensemble of InAs/In$_{0.53}$Al$_{0.24}$Ga$_{0.23}$As/InP quantum dots emitting in the telecom wavelength range. At weak magnetic fields, the major fraction of the total spin polarization decays on the nanosecond time scale due to precession of the individual spins in random nuclear fields. At increased longitudinal magnetic field the spin polarization decays during the sub-microsecond time $T_1$ at low temperatures, which decreases by three orders of magnitude when approaching room temperature. At low temperatures ($T\lesssim 50$~K) we found a relatively strong variation of $T_1$ with $T$ which is so far not understood, while at elevated temperatures ($T\gtrsim 50$~K) $T_1$ is dominated by spin-orbit relaxation with emission and absorption of an optical phonon. The transverse spin relaxation time $T_2$ at elevated temperatures is limited by the $T_1$ time.

\section{Acknowledgements}
\begin{acknowledgments}
We are grateful to I.~A.~Yugova and N.~E.~Kopteva for useful discussions, to E.~Kirstein for help with experiments and to M.~Yacob for the help with the sample growth. Financial support from the Russian Foundation for Basic Research (RFBR, Project~No.~15-52-12019) and Deutsche Forschungsgemeinschaft (DFG, Project A1) in the framework of International Collaborative Research Center TRR 160 is acknowledged. The Dortmund and Kassel teams acknowledge the support by the BMBF in the frame of the Project Q.com-H (Contracts No. 16KIS0104K and No. 16KIS0112). The Dortmund team acknowledges also support by the BMBF-project Q.Link.X (Contract No. 16KIS0857). M. Bayer thanks the Ministry of Education and Science of the Russian Federation (Contract No. 14.Z50.31.0021). A.V.M. acknowledges Saint-Petersburg State University for the research grant 11.34.2.2012. P.S.G. acknowledges support by the Russian Foundation for Basic Research (RFBR, Project~No.~18-32-00568).
\end{acknowledgments}


\begin{thebibliography}{54}%
\makeatletter
\providecommand \@ifxundefined [1]{%
 \@ifx{#1\undefined}
}%
\providecommand \@ifnum [1]{%
 \ifnum #1\expandafter \@firstoftwo
 \else \expandafter \@secondoftwo
 \fi
}%
\providecommand \@ifx [1]{%
 \ifx #1\expandafter \@firstoftwo
 \else \expandafter \@secondoftwo
 \fi
}%
\providecommand \natexlab [1]{#1}%
\providecommand \enquote  [1]{``#1''}%
\providecommand \bibnamefont  [1]{#1}%
\providecommand \bibfnamefont [1]{#1}%
\providecommand \citenamefont [1]{#1}%
\providecommand \href@noop [0]{\@secondoftwo}%
\providecommand \href [0]{\begingroup \@sanitize@url \@href}%
\providecommand \@href[1]{\@@startlink{#1}\@@href}%
\providecommand \@@href[1]{\endgroup#1\@@endlink}%
\providecommand \@sanitize@url [0]{\catcode `\\12\catcode `\$12\catcode
  `\&12\catcode `\#12\catcode `\^12\catcode `\_12\catcode `\%12\relax}%
\providecommand \@@startlink[1]{}%
\providecommand \@@endlink[0]{}%
\providecommand \url  [0]{\begingroup\@sanitize@url \@url }%
\providecommand \@url [1]{\endgroup\@href {#1}{\urlprefix }}%
\providecommand \urlprefix  [0]{URL }%
\providecommand \Eprint [0]{\href }%
\providecommand \doibase [0]{http://dx.doi.org/}%
\providecommand \selectlanguage [0]{\@gobble}%
\providecommand \bibinfo  [0]{\@secondoftwo}%
\providecommand \bibfield  [0]{\@secondoftwo}%
\providecommand \translation [1]{[#1]}%
\providecommand \BibitemOpen [0]{}%
\providecommand \bibitemStop [0]{}%
\providecommand \bibitemNoStop [0]{.\EOS\space}%
\providecommand \EOS [0]{\spacefactor3000\relax}%
\providecommand \BibitemShut  [1]{\csname bibitem#1\endcsname}%
\let\auto@bib@innerbib\@empty
\bibitem [{\citenamefont {Loss}\ and\ \citenamefont
  {DiVincenzo}(1998)}]{Loss1998}%
  \BibitemOpen
  \bibfield  {author} {\bibinfo {author} {\bibfnamefont {D.}~\bibnamefont
  {Loss}}\ and\ \bibinfo {author} {\bibfnamefont {D.~P.}\ \bibnamefont
  {DiVincenzo}},\ }\bibfield  {title} {\enquote {\bibinfo {title} {{Quantum
  computation with quantum dots}},}\ }\href {\doibase 10.1103/PhysRevA.57.120}
  {\bibfield  {journal} {\bibinfo  {journal} {Phys. Rev. A}\ }\textbf {\bibinfo
  {volume} {57}},\ \bibinfo {pages} {120} (\bibinfo {year} {1998})}\BibitemShut
  {NoStop}%
\bibitem [{\citenamefont {Fafard}\ \emph {et~al.}(1996)\citenamefont {Fafard},
  \citenamefont {Wasilewski}, \citenamefont {McCaffrey}, \citenamefont
  {Raymond},\ and\ \citenamefont {Charbonneau}}]{Fafard1996}%
  \BibitemOpen
  \bibfield  {author} {\bibinfo {author} {\bibfnamefont {S.}~\bibnamefont
  {Fafard}}, \bibinfo {author} {\bibfnamefont {Z.}~\bibnamefont {Wasilewski}},
  \bibinfo {author} {\bibfnamefont {J.}~\bibnamefont {McCaffrey}}, \bibinfo
  {author} {\bibfnamefont {S.}~\bibnamefont {Raymond}}, \ and\ \bibinfo
  {author} {\bibfnamefont {S.}~\bibnamefont {Charbonneau}},\ }\bibfield
  {title} {\enquote {\bibinfo {title} {{InAs self‐assembled quantum dots on
  InP by molecular beam epitaxy}},}\ }\href {\doibase 10.1063/1.116122}
  {\bibfield  {journal} {\bibinfo  {journal} {Appl. Phys. Lett.}\ }\textbf
  {\bibinfo {volume} {68}},\ \bibinfo {pages} {991} (\bibinfo {year}
  {1996})}\BibitemShut {NoStop}%
\bibitem [{\citenamefont {Ponchet}\ \emph {et~al.}(1995)\citenamefont
  {Ponchet}, \citenamefont {{Le Corre}}, \citenamefont {L'Haridon},
  \citenamefont {Lambert},\ and\ \citenamefont {Sala{\"{u}}n}}]{Ponchet1995}%
  \BibitemOpen
  \bibfield  {author} {\bibinfo {author} {\bibfnamefont {A.}~\bibnamefont
  {Ponchet}}, \bibinfo {author} {\bibfnamefont {A.}~\bibnamefont {{Le Corre}}},
  \bibinfo {author} {\bibfnamefont {H.}~\bibnamefont {L'Haridon}}, \bibinfo
  {author} {\bibfnamefont {B.}~\bibnamefont {Lambert}}, \ and\ \bibinfo
  {author} {\bibfnamefont {S.}~\bibnamefont {Sala{\"{u}}n}},\ }\bibfield
  {title} {\enquote {\bibinfo {title} {{Relationship between
  self‐organization and size of InAs islands on InP(001) grown by
  gas‐source molecular beam epitaxy}},}\ }\href {\doibase 10.1063/1.114353}
  {\bibfield  {journal} {\bibinfo  {journal} {Appl. Phys. Lett.}\ }\textbf
  {\bibinfo {volume} {67}},\ \bibinfo {pages} {1850} (\bibinfo {year}
  {1995})}\BibitemShut {NoStop}%
\bibitem [{\citenamefont {Taskinen}\ \emph {et~al.}(1997)\citenamefont
  {Taskinen}, \citenamefont {Sopanen}, \citenamefont {Lipsanen}, \citenamefont
  {Tulkki}, \citenamefont {Tuomi},\ and\ \citenamefont
  {Ahopelto}}]{Taskinen1997}%
  \BibitemOpen
  \bibfield  {author} {\bibinfo {author} {\bibfnamefont {M.}~\bibnamefont
  {Taskinen}}, \bibinfo {author} {\bibfnamefont {M.}~\bibnamefont {Sopanen}},
  \bibinfo {author} {\bibfnamefont {H.}~\bibnamefont {Lipsanen}}, \bibinfo
  {author} {\bibfnamefont {J.}~\bibnamefont {Tulkki}}, \bibinfo {author}
  {\bibfnamefont {T.}~\bibnamefont {Tuomi}}, \ and\ \bibinfo {author}
  {\bibfnamefont {J.}~\bibnamefont {Ahopelto}},\ }\bibfield  {title} {\enquote
  {\bibinfo {title} {{Self-organized InAs islands on (100) InP by metalorganic
  vapor-phase epitaxy}},}\ }\href {\doibase 10.1016/S0039-6028(96)01597-X}
  {\bibfield  {journal} {\bibinfo  {journal} {Surf. Sci.}\ }\textbf {\bibinfo
  {volume} {376}},\ \bibinfo {pages} {60} (\bibinfo {year} {1997})}\BibitemShut
  {NoStop}%
\bibitem [{\citenamefont {Paranthoen}\ \emph {et~al.}(2001)\citenamefont
  {Paranthoen}, \citenamefont {Bertru}, \citenamefont {Dehaese}, \citenamefont
  {{Le Corre}}, \citenamefont {Loualiche}, \citenamefont {Lambert},\ and\
  \citenamefont {Patriarche}}]{Paranthoen2001}%
  \BibitemOpen
  \bibfield  {author} {\bibinfo {author} {\bibfnamefont {C.}~\bibnamefont
  {Paranthoen}}, \bibinfo {author} {\bibfnamefont {N.}~\bibnamefont {Bertru}},
  \bibinfo {author} {\bibfnamefont {O.}~\bibnamefont {Dehaese}}, \bibinfo
  {author} {\bibfnamefont {A.}~\bibnamefont {{Le Corre}}}, \bibinfo {author}
  {\bibfnamefont {S.}~\bibnamefont {Loualiche}}, \bibinfo {author}
  {\bibfnamefont {B.}~\bibnamefont {Lambert}}, \ and\ \bibinfo {author}
  {\bibfnamefont {G.}~\bibnamefont {Patriarche}},\ }\bibfield  {title}
  {\enquote {\bibinfo {title} {{Height dispersion control of InAs/InP quantum
  dots emitting at 1.55 $\mu$m}},}\ }\href {\doibase 10.1063/1.1356449}
  {\bibfield  {journal} {\bibinfo  {journal} {Appl. Phys. Lett.}\ }\textbf
  {\bibinfo {volume} {78}},\ \bibinfo {pages} {1751} (\bibinfo {year}
  {2001})}\BibitemShut {NoStop}%
\bibitem [{\citenamefont {Saito}\ \emph {et~al.}(2001)\citenamefont {Saito},
  \citenamefont {Nishi},\ and\ \citenamefont {Sugou}}]{Saito2001}%
  \BibitemOpen
  \bibfield  {author} {\bibinfo {author} {\bibfnamefont {H.}~\bibnamefont
  {Saito}}, \bibinfo {author} {\bibfnamefont {K.}~\bibnamefont {Nishi}}, \ and\
  \bibinfo {author} {\bibfnamefont {S.}~\bibnamefont {Sugou}},\ }\bibfield
  {title} {\enquote {\bibinfo {title} {{Ground-state lasing at room temperature
  in long-wavelength InAs quantum-dot lasers on InP(311)B substrates}},}\
  }\href {\doibase 10.1063/1.1339846} {\bibfield  {journal} {\bibinfo
  {journal} {Appl. Phys. Lett.}\ }\textbf {\bibinfo {volume} {78}},\ \bibinfo
  {pages} {267} (\bibinfo {year} {2001})}\BibitemShut {NoStop}%
\bibitem [{\citenamefont {Wang}\ \emph {et~al.}(2001)\citenamefont {Wang},
  \citenamefont {Stintz}, \citenamefont {Varangis}, \citenamefont {Newell},
  \citenamefont {Li}, \citenamefont {Malloy},\ and\ \citenamefont
  {Lester}}]{Wang2001}%
  \BibitemOpen
  \bibfield  {author} {\bibinfo {author} {\bibfnamefont {R.H.}\ \bibnamefont
  {Wang}}, \bibinfo {author} {\bibfnamefont {A.}~\bibnamefont {Stintz}},
  \bibinfo {author} {\bibfnamefont {P.M.}\ \bibnamefont {Varangis}}, \bibinfo
  {author} {\bibfnamefont {T.C.}\ \bibnamefont {Newell}}, \bibinfo {author}
  {\bibfnamefont {H.}~\bibnamefont {Li}}, \bibinfo {author} {\bibfnamefont
  {K.J.}\ \bibnamefont {Malloy}}, \ and\ \bibinfo {author} {\bibfnamefont
  {L.F.}\ \bibnamefont {Lester}},\ }\bibfield  {title} {\enquote {\bibinfo
  {title} {{Room-temperature operation of InAs quantum-dash lasers on InP
  [001]}},}\ }\href {\doibase 10.1109/68.935797} {\bibfield  {journal}
  {\bibinfo  {journal} {IEEE Photonics Technol. Lett.}\ }\textbf {\bibinfo
  {volume} {13}},\ \bibinfo {pages} {767} (\bibinfo {year} {2001})}\BibitemShut
  {NoStop}%
\bibitem [{\citenamefont {Lelarge}\ \emph {et~al.}(2005)\citenamefont
  {Lelarge}, \citenamefont {Rousseau}, \citenamefont {Dagens}, \citenamefont
  {Poingt}, \citenamefont {Pommereau},\ and\ \citenamefont
  {Accard}}]{Lelarge2005}%
  \BibitemOpen
  \bibfield  {author} {\bibinfo {author} {\bibfnamefont {F.}~\bibnamefont
  {Lelarge}}, \bibinfo {author} {\bibfnamefont {B.}~\bibnamefont {Rousseau}},
  \bibinfo {author} {\bibfnamefont {B.}~\bibnamefont {Dagens}}, \bibinfo
  {author} {\bibfnamefont {F.}~\bibnamefont {Poingt}}, \bibinfo {author}
  {\bibfnamefont {F.}~\bibnamefont {Pommereau}}, \ and\ \bibinfo {author}
  {\bibfnamefont {A.}~\bibnamefont {Accard}},\ }\bibfield  {title} {\enquote
  {\bibinfo {title} {{Room temperature continuous-wave operation of buried
  ridge stripe lasers using InAs-InP (100) quantum dots as active core}},}\
  }\href {\doibase 10.1109/LPT.2005.848279} {\bibfield  {journal} {\bibinfo
  {journal} {IEEE Photonics Technol. Lett.}\ }\textbf {\bibinfo {volume}
  {17}},\ \bibinfo {pages} {1369} (\bibinfo {year} {2005})}\BibitemShut
  {NoStop}%
\bibitem [{\citenamefont {van Veldhoven}\ \emph {et~al.}(2009)\citenamefont
  {van Veldhoven}, \citenamefont {Chauvin}, \citenamefont {Fiore},\ and\
  \citenamefont {N{\"{o}}tzel}}]{VanVeldhoven2009}%
  \BibitemOpen
  \bibfield  {author} {\bibinfo {author} {\bibfnamefont {P.~J.}\ \bibnamefont
  {van Veldhoven}}, \bibinfo {author} {\bibfnamefont {N.}~\bibnamefont
  {Chauvin}}, \bibinfo {author} {\bibfnamefont {A.}~\bibnamefont {Fiore}}, \
  and\ \bibinfo {author} {\bibfnamefont {R.}~\bibnamefont {N{\"{o}}tzel}},\
  }\bibfield  {title} {\enquote {\bibinfo {title} {{Low density 1.55 $\mu$m
  InAs/InGaAsP/InP (100) quantum dots enabled by an ultrathin GaAs
  interlayer}},}\ }\href {\doibase 10.1063/1.3230496} {\bibfield  {journal}
  {\bibinfo  {journal} {Appl. Phys. Lett.}\ }\textbf {\bibinfo {volume} {95}},\
  \bibinfo {pages} {113110} (\bibinfo {year} {2009})}\BibitemShut {NoStop}%
\bibitem [{\citenamefont {Liao}\ \emph {et~al.}(2017)\citenamefont {Liao},
  \citenamefont {Yong}, \citenamefont {Liu}, \citenamefont {Shentu},
  \citenamefont {Li}, \citenamefont {Lin}, \citenamefont {Dai}, \citenamefont
  {Zhao}, \citenamefont {Li}, \citenamefont {Guan}, \citenamefont {Chen},
  \citenamefont {Gong}, \citenamefont {Li}, \citenamefont {Lin}, \citenamefont
  {Pan}, \citenamefont {Pelc}, \citenamefont {Fejer}, \citenamefont {Zhang},
  \citenamefont {Liu}, \citenamefont {Yin}, \citenamefont {Ren}, \citenamefont
  {Wang}, \citenamefont {Zhang}, \citenamefont {Peng},\ and\ \citenamefont
  {Pan}}]{Liao2017}%
  \BibitemOpen
  \bibfield  {author} {\bibinfo {author} {\bibfnamefont {S.-K.}\ \bibnamefont
  {Liao}}, \bibinfo {author} {\bibfnamefont {H.-L.}\ \bibnamefont {Yong}},
  \bibinfo {author} {\bibfnamefont {C.}~\bibnamefont {Liu}}, \bibinfo {author}
  {\bibfnamefont {G.-L.}\ \bibnamefont {Shentu}}, \bibinfo {author}
  {\bibfnamefont {D.-D.}\ \bibnamefont {Li}}, \bibinfo {author} {\bibfnamefont
  {J.}~\bibnamefont {Lin}}, \bibinfo {author} {\bibfnamefont {H.}~\bibnamefont
  {Dai}}, \bibinfo {author} {\bibfnamefont {S.-Q.}\ \bibnamefont {Zhao}},
  \bibinfo {author} {\bibfnamefont {B.}~\bibnamefont {Li}}, \bibinfo {author}
  {\bibfnamefont {J.-Y.}\ \bibnamefont {Guan}}, \bibinfo {author}
  {\bibfnamefont {W.}~\bibnamefont {Chen}}, \bibinfo {author} {\bibfnamefont
  {Y.-H.}\ \bibnamefont {Gong}}, \bibinfo {author} {\bibfnamefont
  {Y.}~\bibnamefont {Li}}, \bibinfo {author} {\bibfnamefont {Z.-H.}\
  \bibnamefont {Lin}}, \bibinfo {author} {\bibfnamefont {G.-S.}\ \bibnamefont
  {Pan}}, \bibinfo {author} {\bibfnamefont {J.~S.}\ \bibnamefont {Pelc}},
  \bibinfo {author} {\bibfnamefont {M.~M.}\ \bibnamefont {Fejer}}, \bibinfo
  {author} {\bibfnamefont {W.-Z.}\ \bibnamefont {Zhang}}, \bibinfo {author}
  {\bibfnamefont {W.-Y.}\ \bibnamefont {Liu}}, \bibinfo {author} {\bibfnamefont
  {J.}~\bibnamefont {Yin}}, \bibinfo {author} {\bibfnamefont {J.-G.}\
  \bibnamefont {Ren}}, \bibinfo {author} {\bibfnamefont {X.-B.}\ \bibnamefont
  {Wang}}, \bibinfo {author} {\bibfnamefont {Q.}~\bibnamefont {Zhang}},
  \bibinfo {author} {\bibfnamefont {C.-Z.}\ \bibnamefont {Peng}}, \ and\
  \bibinfo {author} {\bibfnamefont {J.-W.}\ \bibnamefont {Pan}},\ }\bibfield
  {title} {\enquote {\bibinfo {title} {{Long-distance free-space quantum key
  distribution in daylight towards inter-satellite communication}},}\ }\href
  {\doibase 10.1038/nphoton.2017.116} {\bibfield  {journal} {\bibinfo
  {journal} {Nat. Photonics}\ }\textbf {\bibinfo {volume} {11}},\ \bibinfo
  {pages} {509} (\bibinfo {year} {2017})}\BibitemShut {NoStop}%
\bibitem [{\citenamefont {Sapienza}\ \emph {et~al.}(2016)\citenamefont
  {Sapienza}, \citenamefont {Al-Khuzheyri}, \citenamefont {Dada}, \citenamefont
  {Griffiths}, \citenamefont {Clarke},\ and\ \citenamefont
  {Gerardot}}]{Sapienza2016}%
  \BibitemOpen
  \bibfield  {author} {\bibinfo {author} {\bibfnamefont {L.}~\bibnamefont
  {Sapienza}}, \bibinfo {author} {\bibfnamefont {R.}~\bibnamefont
  {Al-Khuzheyri}}, \bibinfo {author} {\bibfnamefont {A.}~\bibnamefont {Dada}},
  \bibinfo {author} {\bibfnamefont {A.}~\bibnamefont {Griffiths}}, \bibinfo
  {author} {\bibfnamefont {E.}~\bibnamefont {Clarke}}, \ and\ \bibinfo {author}
  {\bibfnamefont {B.~D.}\ \bibnamefont {Gerardot}},\ }\bibfield  {title}
  {\enquote {\bibinfo {title} {{Magneto-optical spectroscopy of single
  charge-tunable InAs/GaAs quantum dots emitting at telecom wavelengths}},}\
  }\href {\doibase 10.1103/PhysRevB.93.155301} {\bibfield  {journal} {\bibinfo
  {journal} {Phys. Rev. B}\ }\textbf {\bibinfo {volume} {93}},\ \bibinfo
  {pages} {155301} (\bibinfo {year} {2016})}\BibitemShut {NoStop}%
\bibitem [{\citenamefont {Mrowinski}\ \emph {et~al.}(2016)\citenamefont
  {Mrowinski}, \citenamefont {Zielinski}, \citenamefont {Swiderski},
  \citenamefont {Misiewicz}, \citenamefont {Somers}, \citenamefont
  {Reithmaier}, \citenamefont {H{\"{o}}fling},\ and\ \citenamefont
  {Sek}}]{Mrowinski2016}%
  \BibitemOpen
  \bibfield  {author} {\bibinfo {author} {\bibfnamefont {P.}~\bibnamefont
  {Mrowinski}}, \bibinfo {author} {\bibfnamefont {M.}~\bibnamefont
  {Zielinski}}, \bibinfo {author} {\bibfnamefont {M.}~\bibnamefont
  {Swiderski}}, \bibinfo {author} {\bibfnamefont {J.}~\bibnamefont
  {Misiewicz}}, \bibinfo {author} {\bibfnamefont {A.}~\bibnamefont {Somers}},
  \bibinfo {author} {\bibfnamefont {J.~P.}\ \bibnamefont {Reithmaier}},
  \bibinfo {author} {\bibfnamefont {S.}~\bibnamefont {H{\"{o}}fling}}, \ and\
  \bibinfo {author} {\bibfnamefont {G.}~\bibnamefont {Sek}},\ }\bibfield
  {title} {\enquote {\bibinfo {title} {{Excitonic fine structure and binding
  energies of excitonic complexes in single InAs quantum dashes}},}\ }\href
  {\doibase 10.1103/PhysRevB.94.115434} {\bibfield  {journal} {\bibinfo
  {journal} {Phys. Rev. B}\ }\textbf {\bibinfo {volume} {94}},\ \bibinfo
  {pages} {115434} (\bibinfo {year} {2016})}\BibitemShut {NoStop}%
\bibitem [{\citenamefont {Miyazawa}\ \emph {et~al.}(2005)\citenamefont
  {Miyazawa}, \citenamefont {Takemoto}, \citenamefont {Sakuma}, \citenamefont
  {Hirose}, \citenamefont {Usuki}, \citenamefont {Yokoyama}, \citenamefont
  {Takatsu},\ and\ \citenamefont {Arakawa}}]{Miyazawa2005}%
  \BibitemOpen
  \bibfield  {author} {\bibinfo {author} {\bibfnamefont {T.}~\bibnamefont
  {Miyazawa}}, \bibinfo {author} {\bibfnamefont {K.}~\bibnamefont {Takemoto}},
  \bibinfo {author} {\bibfnamefont {Y.}~\bibnamefont {Sakuma}}, \bibinfo
  {author} {\bibfnamefont {S.}~\bibnamefont {Hirose}}, \bibinfo {author}
  {\bibfnamefont {T.}~\bibnamefont {Usuki}}, \bibinfo {author} {\bibfnamefont
  {N.}~\bibnamefont {Yokoyama}}, \bibinfo {author} {\bibfnamefont
  {M.}~\bibnamefont {Takatsu}}, \ and\ \bibinfo {author} {\bibfnamefont
  {Y.}~\bibnamefont {Arakawa}},\ }\bibfield  {title} {\enquote {\bibinfo
  {title} {{Single-Photon Generation in the 1.55-µm Optical-Fiber Band from an
  InAs/InP Quantum Dot}},}\ }\href {\doibase 10.1143/JJAP.44.L620} {\bibfield
  {journal} {\bibinfo  {journal} {Jpn. J. Appl. Phys.}\ }\textbf {\bibinfo
  {volume} {44}},\ \bibinfo {pages} {L620} (\bibinfo {year}
  {2005})}\BibitemShut {NoStop}%
\bibitem [{\citenamefont {Takemoto}\ \emph {et~al.}(2007)\citenamefont
  {Takemoto}, \citenamefont {Takatsu}, \citenamefont {Hirose}, \citenamefont
  {Yokoyama}, \citenamefont {Sakuma}, \citenamefont {Usuki}, \citenamefont
  {Miyazawa},\ and\ \citenamefont {Arakawa}}]{Takemoto2007}%
  \BibitemOpen
  \bibfield  {author} {\bibinfo {author} {\bibfnamefont {K.}~\bibnamefont
  {Takemoto}}, \bibinfo {author} {\bibfnamefont {M.}~\bibnamefont {Takatsu}},
  \bibinfo {author} {\bibfnamefont {S.}~\bibnamefont {Hirose}}, \bibinfo
  {author} {\bibfnamefont {N.}~\bibnamefont {Yokoyama}}, \bibinfo {author}
  {\bibfnamefont {Y.}~\bibnamefont {Sakuma}}, \bibinfo {author} {\bibfnamefont
  {T.}~\bibnamefont {Usuki}}, \bibinfo {author} {\bibfnamefont
  {T.}~\bibnamefont {Miyazawa}}, \ and\ \bibinfo {author} {\bibfnamefont
  {Y.}~\bibnamefont {Arakawa}},\ }\bibfield  {title} {\enquote {\bibinfo
  {title} {{An optical horn structure for single-photon source using quantum
  dots at telecommunication wavelength}},}\ }\href {\doibase 10.1063/1.2723177}
  {\bibfield  {journal} {\bibinfo  {journal} {J. Appl. Phys.}\ }\textbf
  {\bibinfo {volume} {101}},\ \bibinfo {pages} {081720} (\bibinfo {year}
  {2007})}\BibitemShut {NoStop}%
\bibitem [{\citenamefont {Takemoto}\ \emph {et~al.}(2010)\citenamefont
  {Takemoto}, \citenamefont {Nambu}, \citenamefont {Miyazawa}, \citenamefont
  {Wakui}, \citenamefont {Hirose}, \citenamefont {Usuki}, \citenamefont
  {Takatsu}, \citenamefont {Yokoyama}, \citenamefont {Yoshino}, \citenamefont
  {Tomita}, \citenamefont {Yorozu}, \citenamefont {Sakuma},\ and\ \citenamefont
  {Arakawa}}]{Takemoto2010}%
  \BibitemOpen
  \bibfield  {author} {\bibinfo {author} {\bibfnamefont {K.}~\bibnamefont
  {Takemoto}}, \bibinfo {author} {\bibfnamefont {Y.}~\bibnamefont {Nambu}},
  \bibinfo {author} {\bibfnamefont {T.}~\bibnamefont {Miyazawa}}, \bibinfo
  {author} {\bibfnamefont {K.}~\bibnamefont {Wakui}}, \bibinfo {author}
  {\bibfnamefont {S.}~\bibnamefont {Hirose}}, \bibinfo {author} {\bibfnamefont
  {T.}~\bibnamefont {Usuki}}, \bibinfo {author} {\bibfnamefont
  {M.}~\bibnamefont {Takatsu}}, \bibinfo {author} {\bibfnamefont
  {N.}~\bibnamefont {Yokoyama}}, \bibinfo {author} {\bibfnamefont
  {K.}~\bibnamefont {Yoshino}}, \bibinfo {author} {\bibfnamefont
  {A.}~\bibnamefont {Tomita}}, \bibinfo {author} {\bibfnamefont
  {S.}~\bibnamefont {Yorozu}}, \bibinfo {author} {\bibfnamefont
  {Y.}~\bibnamefont {Sakuma}}, \ and\ \bibinfo {author} {\bibfnamefont
  {Y.}~\bibnamefont {Arakawa}},\ }\bibfield  {title} {\enquote {\bibinfo
  {title} {{Transmission Experiment of Quantum Keys over 50 km Using
  High-Performance Quantum-Dot Single-Photon Source at 1.5 µm Wavelength}},}\
  }\href {\doibase 10.1143/APEX.3.092802} {\bibfield  {journal} {\bibinfo
  {journal} {Appl. Phys. Express}\ }\textbf {\bibinfo {volume} {3}},\ \bibinfo
  {pages} {092802} (\bibinfo {year} {2010})}\BibitemShut {NoStop}%
\bibitem [{\citenamefont {Birowosuto}\ \emph {et~al.}(2012)\citenamefont
  {Birowosuto}, \citenamefont {Sumikura}, \citenamefont {Matsuo}, \citenamefont
  {Taniyama}, \citenamefont {van Veldhoven}, \citenamefont {N{\"{o}}tzel},\
  and\ \citenamefont {Notomi}}]{Birowosuto2012}%
  \BibitemOpen
  \bibfield  {author} {\bibinfo {author} {\bibfnamefont {M.~D.}\ \bibnamefont
  {Birowosuto}}, \bibinfo {author} {\bibfnamefont {H.}~\bibnamefont
  {Sumikura}}, \bibinfo {author} {\bibfnamefont {S.}~\bibnamefont {Matsuo}},
  \bibinfo {author} {\bibfnamefont {H.}~\bibnamefont {Taniyama}}, \bibinfo
  {author} {\bibfnamefont {P.~J.}\ \bibnamefont {van Veldhoven}}, \bibinfo
  {author} {\bibfnamefont {R.}~\bibnamefont {N{\"{o}}tzel}}, \ and\ \bibinfo
  {author} {\bibfnamefont {M.}~\bibnamefont {Notomi}},\ }\bibfield  {title}
  {\enquote {\bibinfo {title} {{Fast Purcell-enhanced single photon source in
  1,550-nm telecom band from a resonant quantum dot-cavity coupling}},}\ }\href
  {\doibase 10.1038/srep00321} {\bibfield  {journal} {\bibinfo  {journal} {Sci.
  Rep.}\ }\textbf {\bibinfo {volume} {2}},\ \bibinfo {pages} {321} (\bibinfo
  {year} {2012})}\BibitemShut {NoStop}%
\bibitem [{\citenamefont {Benyoucef}\ \emph {et~al.}(2013)\citenamefont
  {Benyoucef}, \citenamefont {Yacob}, \citenamefont {Reithmaier}, \citenamefont
  {Kettler},\ and\ \citenamefont {Michler}}]{Benyoucef2013}%
  \BibitemOpen
  \bibfield  {author} {\bibinfo {author} {\bibfnamefont {M.}~\bibnamefont
  {Benyoucef}}, \bibinfo {author} {\bibfnamefont {M.}~\bibnamefont {Yacob}},
  \bibinfo {author} {\bibfnamefont {J.~P.}\ \bibnamefont {Reithmaier}},
  \bibinfo {author} {\bibfnamefont {J.}~\bibnamefont {Kettler}}, \ and\
  \bibinfo {author} {\bibfnamefont {P.}~\bibnamefont {Michler}},\ }\bibfield
  {title} {\enquote {\bibinfo {title} {{Telecom-wavelength (1.5 $\mu$ m)
  single-photon emission from InP-based quantum dots}},}\ }\href {\doibase
  10.1063/1.4825106} {\bibfield  {journal} {\bibinfo  {journal} {Appl. Phys.
  Lett.}\ }\textbf {\bibinfo {volume} {103}},\ \bibinfo {pages} {162101}
  (\bibinfo {year} {2013})}\BibitemShut {NoStop}%
\bibitem [{\citenamefont {Liu}\ \emph {et~al.}(2013)\citenamefont {Liu},
  \citenamefont {Akahane}, \citenamefont {Jahan}, \citenamefont {Kobayashi},
  \citenamefont {Sasaki}, \citenamefont {Kumano},\ and\ \citenamefont
  {Suemune}}]{Liu2013}%
  \BibitemOpen
  \bibfield  {author} {\bibinfo {author} {\bibfnamefont {X.}~\bibnamefont
  {Liu}}, \bibinfo {author} {\bibfnamefont {K.}~\bibnamefont {Akahane}},
  \bibinfo {author} {\bibfnamefont {N.~A.}\ \bibnamefont {Jahan}}, \bibinfo
  {author} {\bibfnamefont {N.}~\bibnamefont {Kobayashi}}, \bibinfo {author}
  {\bibfnamefont {M.}~\bibnamefont {Sasaki}}, \bibinfo {author} {\bibfnamefont
  {H.}~\bibnamefont {Kumano}}, \ and\ \bibinfo {author} {\bibfnamefont
  {I.}~\bibnamefont {Suemune}},\ }\bibfield  {title} {\enquote {\bibinfo
  {title} {{Single-photon emission in telecommunication band from an InAs
  quantum dot grown on InP with molecular-beam epitaxy}},}\ }\href {\doibase
  10.1063/1.4817940} {\bibfield  {journal} {\bibinfo  {journal} {Appl. Phys.
  Lett.}\ }\textbf {\bibinfo {volume} {103}},\ \bibinfo {pages} {061114}
  (\bibinfo {year} {2013})}\BibitemShut {NoStop}%
\bibitem [{\citenamefont {Dusanowski}\ \emph {et~al.}(2014)\citenamefont
  {Dusanowski}, \citenamefont {Syperek}, \citenamefont {Mrowi{\'{n}}ski},
  \citenamefont {Rudno-Rudzi{\'{n}}ski}, \citenamefont {Misiewicz},
  \citenamefont {Somers}, \citenamefont {H{\"{o}}fling}, \citenamefont {Kamp},
  \citenamefont {Reithmaier},\ and\ \citenamefont
  {S{\c{e}}k}}]{Dusanowski2014}%
  \BibitemOpen
  \bibfield  {author} {\bibinfo {author} {\bibfnamefont {L.}~\bibnamefont
  {Dusanowski}}, \bibinfo {author} {\bibfnamefont {M.}~\bibnamefont {Syperek}},
  \bibinfo {author} {\bibfnamefont {P.}~\bibnamefont {Mrowi{\'{n}}ski}},
  \bibinfo {author} {\bibfnamefont {W.}~\bibnamefont {Rudno-Rudzi{\'{n}}ski}},
  \bibinfo {author} {\bibfnamefont {J.}~\bibnamefont {Misiewicz}}, \bibinfo
  {author} {\bibfnamefont {A.}~\bibnamefont {Somers}}, \bibinfo {author}
  {\bibfnamefont {S.}~\bibnamefont {H{\"{o}}fling}}, \bibinfo {author}
  {\bibfnamefont {M.}~\bibnamefont {Kamp}}, \bibinfo {author} {\bibfnamefont
  {J.~P.}\ \bibnamefont {Reithmaier}}, \ and\ \bibinfo {author} {\bibfnamefont
  {G.}~\bibnamefont {S{\c{e}}k}},\ }\bibfield  {title} {\enquote {\bibinfo
  {title} {{Single photon emission at 1.55 $\mu$m from charged and neutral
  exciton confined in a single quantum dash}},}\ }\href {\doibase
  10.1063/1.4890603} {\bibfield  {journal} {\bibinfo  {journal} {Appl. Phys.
  Lett.}\ }\textbf {\bibinfo {volume} {105}},\ \bibinfo {pages} {021909}
  (\bibinfo {year} {2014})}\BibitemShut {NoStop}%
\bibitem [{\citenamefont {Kim}\ \emph {et~al.}(2016)\citenamefont {Kim},
  \citenamefont {Cai}, \citenamefont {Richardson}, \citenamefont {Leavitt},\
  and\ \citenamefont {Waks}}]{Kim2016}%
  \BibitemOpen
  \bibfield  {author} {\bibinfo {author} {\bibfnamefont {J.-H.}\ \bibnamefont
  {Kim}}, \bibinfo {author} {\bibfnamefont {T.}~\bibnamefont {Cai}}, \bibinfo
  {author} {\bibfnamefont {C.~J.~K.}\ \bibnamefont {Richardson}}, \bibinfo
  {author} {\bibfnamefont {R.~P.}\ \bibnamefont {Leavitt}}, \ and\ \bibinfo
  {author} {\bibfnamefont {E.}~\bibnamefont {Waks}},\ }\bibfield  {title}
  {\enquote {\bibinfo {title} {{Two-photon interference from a bright
  single-photon source at telecom wavelengths}},}\ }\href {\doibase
  10.1364/OPTICA.3.000577} {\bibfield  {journal} {\bibinfo  {journal} {Optica}\
  }\textbf {\bibinfo {volume} {3}},\ \bibinfo {pages} {577} (\bibinfo {year}
  {2016})}\BibitemShut {NoStop}%
\bibitem [{\citenamefont {Paul}\ \emph {et~al.}(2017)\citenamefont {Paul},
  \citenamefont {Olbrich}, \citenamefont {H{\"{o}}schele}, \citenamefont
  {Schreier}, \citenamefont {Kettler}, \citenamefont {Portalupi}, \citenamefont
  {Jetter},\ and\ \citenamefont {Michler}}]{Paul2017}%
  \BibitemOpen
  \bibfield  {author} {\bibinfo {author} {\bibfnamefont {M.}~\bibnamefont
  {Paul}}, \bibinfo {author} {\bibfnamefont {F.}~\bibnamefont {Olbrich}},
  \bibinfo {author} {\bibfnamefont {J.}~\bibnamefont {H{\"{o}}schele}},
  \bibinfo {author} {\bibfnamefont {S.}~\bibnamefont {Schreier}}, \bibinfo
  {author} {\bibfnamefont {J.}~\bibnamefont {Kettler}}, \bibinfo {author}
  {\bibfnamefont {S.~L.}\ \bibnamefont {Portalupi}}, \bibinfo {author}
  {\bibfnamefont {M.}~\bibnamefont {Jetter}}, \ and\ \bibinfo {author}
  {\bibfnamefont {P.}~\bibnamefont {Michler}},\ }\bibfield  {title} {\enquote
  {\bibinfo {title} {{Single-photon emission at 1.55 $\mu$ m from MOVPE-grown
  InAs quantum dots on InGaAs/GaAs metamorphic buffers}},}\ }\href {\doibase
  10.1063/1.4993935} {\bibfield  {journal} {\bibinfo  {journal} {Appl. Phys.
  Lett.}\ }\textbf {\bibinfo {volume} {111}},\ \bibinfo {pages} {033102}
  (\bibinfo {year} {2017})}\BibitemShut {NoStop}%
\bibitem [{\citenamefont {Olbrich}\ \emph {et~al.}(2017)\citenamefont
  {Olbrich}, \citenamefont {H{\"{o}}schele}, \citenamefont {M{\"{u}}ller},
  \citenamefont {Kettler}, \citenamefont {Portalupi}, \citenamefont {Paul},
  \citenamefont {Jetter},\ and\ \citenamefont {Michler}}]{Olbrich2017}%
  \BibitemOpen
  \bibfield  {author} {\bibinfo {author} {\bibfnamefont {F.}~\bibnamefont
  {Olbrich}}, \bibinfo {author} {\bibfnamefont {J.}~\bibnamefont
  {H{\"{o}}schele}}, \bibinfo {author} {\bibfnamefont {M.}~\bibnamefont
  {M{\"{u}}ller}}, \bibinfo {author} {\bibfnamefont {J.}~\bibnamefont
  {Kettler}}, \bibinfo {author} {\bibfnamefont {S.~L.}\ \bibnamefont
  {Portalupi}}, \bibinfo {author} {\bibfnamefont {M.}~\bibnamefont {Paul}},
  \bibinfo {author} {\bibfnamefont {M.}~\bibnamefont {Jetter}}, \ and\ \bibinfo
  {author} {\bibfnamefont {P.}~\bibnamefont {Michler}},\ }\bibfield  {title}
  {\enquote {\bibinfo {title} {{Polarization-entangled photons from an
  InGaAs-based quantum dot emitting in the telecom C-band}},}\ }\href {\doibase
  10.1063/1.4994145} {\bibfield  {journal} {\bibinfo  {journal} {Appl. Phys.
  Lett.}\ }\textbf {\bibinfo {volume} {111}},\ \bibinfo {pages} {133106}
  (\bibinfo {year} {2017})}\BibitemShut {NoStop}%
\bibitem [{\citenamefont {Yakovlev}\ and\ \citenamefont
  {Bayer}(2017)}]{Dyakonov2017}%
  \BibitemOpen
  \bibfield  {author} {\bibinfo {author} {\bibfnamefont {D.R.}\ \bibnamefont
  {Yakovlev}}\ and\ \bibinfo {author} {\bibfnamefont {M.}~\bibnamefont
  {Bayer}},\ }\bibfield  {title} {\enquote {\bibinfo {title} {{Coherent Spin
  Dynamics of Carriers}},}\ }in\ \href {\doibase 10.1007/978-3-319-65436-2_6}
  {\emph {\bibinfo {booktitle} {Spin Physics in Semiconductors}}},\ \bibinfo {series}
  {Springer Series in Solid-State Sciences}, Vol.\ \bibinfo {volume} {157},\
  \bibinfo {editor} {edited by\ \bibinfo {editor} {\bibfnamefont {M.~I.}\
  \bibnamefont {Dyakonov}}}\ (\bibinfo  {publisher} {Springer International
  Publishing},\ \bibinfo {address} {Cham},\ \bibinfo {year} {2017})\ p.\
  \bibinfo {pages} {155}\BibitemShut {NoStop}%
\bibitem [{\citenamefont {Bayer}\ \emph {et~al.}(1999)\citenamefont {Bayer},
  \citenamefont {Kuther}, \citenamefont {Forchel}, \citenamefont {Gorbunov},
  \citenamefont {Timofeev}, \citenamefont {Sch{\"{a}}fer}, \citenamefont
  {Reithmaier}, \citenamefont {Reinecke},\ and\ \citenamefont
  {Walck}}]{Bayer1999}%
  \BibitemOpen
  \bibfield  {author} {\bibinfo {author} {\bibfnamefont {M.}~\bibnamefont
  {Bayer}}, \bibinfo {author} {\bibfnamefont {A.}~\bibnamefont {Kuther}},
  \bibinfo {author} {\bibfnamefont {A.}~\bibnamefont {Forchel}}, \bibinfo
  {author} {\bibfnamefont {A.}~\bibnamefont {Gorbunov}}, \bibinfo {author}
  {\bibfnamefont {V.~B.}\ \bibnamefont {Timofeev}}, \bibinfo {author}
  {\bibfnamefont {F.}~\bibnamefont {Sch{\"{a}}fer}}, \bibinfo {author}
  {\bibfnamefont {J.~P.}\ \bibnamefont {Reithmaier}}, \bibinfo {author}
  {\bibfnamefont {T.~L.}\ \bibnamefont {Reinecke}}, \ and\ \bibinfo {author}
  {\bibfnamefont {S.~N.}\ \bibnamefont {Walck}},\ }\bibfield  {title} {\enquote
  {\bibinfo {title} {{Electron and Hole g Factors and Exchange Interaction from
  Studies of the Exciton Fine Structure in In0.60Ga0.40As Quantum Dots}},}\
  }\href {\doibase 10.1103/PhysRevLett.82.1748} {\bibfield  {journal} {\bibinfo
   {journal} {Phys. Rev. Lett.}\ }\textbf {\bibinfo {volume} {82}},\ \bibinfo
  {pages} {1748} (\bibinfo {year} {1999})}\BibitemShut {NoStop}%
\bibitem [{\citenamefont {Kroutvar}\ \emph {et~al.}(2004)\citenamefont
  {Kroutvar}, \citenamefont {Ducommun}, \citenamefont {Heiss}, \citenamefont
  {Bichler}, \citenamefont {Schuh}, \citenamefont {Abstreiter},\ and\
  \citenamefont {Finley}}]{Kroutvar2004}%
  \BibitemOpen
  \bibfield  {author} {\bibinfo {author} {\bibfnamefont {M.}~\bibnamefont
  {Kroutvar}}, \bibinfo {author} {\bibfnamefont {Y.}~\bibnamefont {Ducommun}},
  \bibinfo {author} {\bibfnamefont {D.}~\bibnamefont {Heiss}}, \bibinfo
  {author} {\bibfnamefont {M.}~\bibnamefont {Bichler}}, \bibinfo {author}
  {\bibfnamefont {D.}~\bibnamefont {Schuh}}, \bibinfo {author} {\bibfnamefont
  {G.}~\bibnamefont {Abstreiter}}, \ and\ \bibinfo {author} {\bibfnamefont
  {J.~J.}\ \bibnamefont {Finley}},\ }\bibfield  {title} {\enquote {\bibinfo
  {title} {{Optically programmable electron spin memory using semiconductor
  quantum dots}},}\ }\href {\doibase 10.1038/nature03008} {\bibfield  {journal}
  {\bibinfo  {journal} {Nature}\ }\textbf {\bibinfo {volume} {432}},\ \bibinfo
  {pages} {81} (\bibinfo {year} {2004})}\BibitemShut {NoStop}%
\bibitem [{\citenamefont {Dutt}\ \emph {et~al.}(2005)\citenamefont {Dutt},
  \citenamefont {Cheng}, \citenamefont {Li}, \citenamefont {Xu}, \citenamefont
  {Li}, \citenamefont {Berman}, \citenamefont {Steel}, \citenamefont {Bracker},
  \citenamefont {Gammon}, \citenamefont {Economou}, \citenamefont {Liu},\ and\
  \citenamefont {Sham}}]{Dutt2005}%
  \BibitemOpen
  \bibfield  {author} {\bibinfo {author} {\bibfnamefont {M.~V.~G.}\
  \bibnamefont {Dutt}}, \bibinfo {author} {\bibfnamefont {J.}~\bibnamefont
  {Cheng}}, \bibinfo {author} {\bibfnamefont {B.}~\bibnamefont {Li}}, \bibinfo
  {author} {\bibfnamefont {X.}~\bibnamefont {Xu}}, \bibinfo {author}
  {\bibfnamefont {X.}~\bibnamefont {Li}}, \bibinfo {author} {\bibfnamefont
  {P.~R.}\ \bibnamefont {Berman}}, \bibinfo {author} {\bibfnamefont {D.~G.}\
  \bibnamefont {Steel}}, \bibinfo {author} {\bibfnamefont {A.~S.}\ \bibnamefont
  {Bracker}}, \bibinfo {author} {\bibfnamefont {D.}~\bibnamefont {Gammon}},
  \bibinfo {author} {\bibfnamefont {S.~E.}\ \bibnamefont {Economou}}, \bibinfo
  {author} {\bibfnamefont {R.~B.}\ \bibnamefont {Liu}}, \ and\ \bibinfo
  {author} {\bibfnamefont {L.~J.}\ \bibnamefont {Sham}},\ }\bibfield  {title}
  {\enquote {\bibinfo {title} {{Stimulated and spontaneous optical generation
  of electron spin coherence in charged GaAs quantum dots}},}\ }\href {\doibase
  10.1103/PhysRevLett.94.227403} {\bibfield  {journal} {\bibinfo  {journal}
  {Phys. Rev. Lett.}\ }\textbf {\bibinfo {volume} {94}},\ \bibinfo {pages}
  {227403} (\bibinfo {year} {2005})}\BibitemShut {NoStop}%
\bibitem [{\citenamefont {Greilich}\ \emph {et~al.}(2006)\citenamefont
  {Greilich}, \citenamefont {Oulton}, \citenamefont {Zhukov}, \citenamefont
  {Yugova}, \citenamefont {Yakovlev}, \citenamefont {Bayer}, \citenamefont
  {Shabaev}, \citenamefont {Efros}, \citenamefont {Merkulov}, \citenamefont
  {Stavarache}, \citenamefont {Reuter},\ and\ \citenamefont
  {Wieck}}]{Greilich2006}%
  \BibitemOpen
  \bibfield  {author} {\bibinfo {author} {\bibfnamefont {A.}~\bibnamefont
  {Greilich}}, \bibinfo {author} {\bibfnamefont {R.}~\bibnamefont {Oulton}},
  \bibinfo {author} {\bibfnamefont {E.~A.}\ \bibnamefont {Zhukov}}, \bibinfo
  {author} {\bibfnamefont {I.~A.}\ \bibnamefont {Yugova}}, \bibinfo {author}
  {\bibfnamefont {D.~R.}\ \bibnamefont {Yakovlev}}, \bibinfo {author}
  {\bibfnamefont {M.}~\bibnamefont {Bayer}}, \bibinfo {author} {\bibfnamefont
  {A.}~\bibnamefont {Shabaev}}, \bibinfo {author} {\bibfnamefont {Al~L.}\
  \bibnamefont {Efros}}, \bibinfo {author} {\bibfnamefont {I.~A.}\ \bibnamefont
  {Merkulov}}, \bibinfo {author} {\bibfnamefont {V.}~\bibnamefont
  {Stavarache}}, \bibinfo {author} {\bibfnamefont {D.}~\bibnamefont {Reuter}},
  \ and\ \bibinfo {author} {\bibfnamefont {A.}~\bibnamefont {Wieck}},\
  }\bibfield  {title} {\enquote {\bibinfo {title} {{Optical control of spin
  coherence in singly charged (In,Ga)As/GaAs quantum dots}},}\ }\href {\doibase
  10.1103/PhysRevLett.96.227401} {\bibfield  {journal} {\bibinfo  {journal}
  {Phys. Rev. Lett.}\ }\textbf {\bibinfo {volume} {96}},\ \bibinfo {pages}
  {227401} (\bibinfo {year} {2006})}\BibitemShut {NoStop}%
\bibitem [{\citenamefont {Hernandez}\ \emph {et~al.}(2008)\citenamefont
  {Hernandez}, \citenamefont {Greilich}, \citenamefont {Brito}, \citenamefont
  {Wiemann}, \citenamefont {Yakovlev}, \citenamefont {Reuter}, \citenamefont
  {Wieck},\ and\ \citenamefont {Bayer}}]{Hernandez2008}%
  \BibitemOpen
  \bibfield  {author} {\bibinfo {author} {\bibfnamefont {F.~G.~G.}\
  \bibnamefont {Hernandez}}, \bibinfo {author} {\bibfnamefont {A.}~\bibnamefont
  {Greilich}}, \bibinfo {author} {\bibfnamefont {F.}~\bibnamefont {Brito}},
  \bibinfo {author} {\bibfnamefont {M.}~\bibnamefont {Wiemann}}, \bibinfo
  {author} {\bibfnamefont {D.~R.}\ \bibnamefont {Yakovlev}}, \bibinfo {author}
  {\bibfnamefont {D.}~\bibnamefont {Reuter}}, \bibinfo {author} {\bibfnamefont
  {A.~D.}\ \bibnamefont {Wieck}}, \ and\ \bibinfo {author} {\bibfnamefont
  {M.}~\bibnamefont {Bayer}},\ }\bibfield  {title} {\enquote {\bibinfo {title}
  {{Temperature-induced spin-coherence dissipation in quantum dots}},}\ }\href
  {\doibase 10.1103/PhysRevB.78.041303} {\bibfield  {journal} {\bibinfo
  {journal} {Phys. Rev. B}\ }\textbf {\bibinfo {volume} {78}},\ \bibinfo
  {pages} {041303} (\bibinfo {year} {2008})}\BibitemShut {NoStop}%
\bibitem [{\citenamefont {Fras}\ \emph {et~al.}(2012)\citenamefont {Fras},
  \citenamefont {Eble}, \citenamefont {Desfonds}, \citenamefont {Bernardot},
  \citenamefont {Testelin}, \citenamefont {Chamarro}, \citenamefont {Miard},\
  and\ \citenamefont {Lema{\^{i}}tre}}]{Fras2012}%
  \BibitemOpen
  \bibfield  {author} {\bibinfo {author} {\bibfnamefont {F.}~\bibnamefont
  {Fras}}, \bibinfo {author} {\bibfnamefont {B.}~\bibnamefont {Eble}}, \bibinfo
  {author} {\bibfnamefont {P.}~\bibnamefont {Desfonds}}, \bibinfo {author}
  {\bibfnamefont {F.}~\bibnamefont {Bernardot}}, \bibinfo {author}
  {\bibfnamefont {C.}~\bibnamefont {Testelin}}, \bibinfo {author}
  {\bibfnamefont {M.}~\bibnamefont {Chamarro}}, \bibinfo {author}
  {\bibfnamefont {A.}~\bibnamefont {Miard}}, \ and\ \bibinfo {author}
  {\bibfnamefont {A.}~\bibnamefont {Lema{\^{i}}tre}},\ }\bibfield  {title}
  {\enquote {\bibinfo {title} {{Two-phonon process and hyperfine interaction
  limiting slow hole-spin relaxation time in InAs/GaAs quantum dots}},}\ }\href
  {\doibase 10.1103/PhysRevB.86.045306} {\bibfield  {journal} {\bibinfo
  {journal} {Phys. Rev. B}\ }\textbf {\bibinfo {volume} {86}},\ \bibinfo
  {pages} {045306} (\bibinfo {year} {2012})}\BibitemShut {NoStop}%
\bibitem [{\citenamefont {Dou}\ \emph {et~al.}(2012)\citenamefont {Dou},
  \citenamefont {Sun}, \citenamefont {Jiang}, \citenamefont {Ni},\ and\
  \citenamefont {Niu}}]{Dou2012}%
  \BibitemOpen
  \bibfield  {author} {\bibinfo {author} {\bibfnamefont {X.~M.}\ \bibnamefont
  {Dou}}, \bibinfo {author} {\bibfnamefont {B.~Q.}\ \bibnamefont {Sun}},
  \bibinfo {author} {\bibfnamefont {D.~S.}\ \bibnamefont {Jiang}}, \bibinfo
  {author} {\bibfnamefont {H.~Q.}\ \bibnamefont {Ni}}, \ and\ \bibinfo {author}
  {\bibfnamefont {Z.~C.}\ \bibnamefont {Niu}},\ }\bibfield  {title} {\enquote
  {\bibinfo {title} {{Temperature dependence of electron-spin relaxation in a
  single InAs quantum dot at zero applied magnetic field}},}\ }\href {\doibase
  10.1063/1.3692066} {\bibfield  {journal} {\bibinfo  {journal} {J. Appl.
  Phys.}\ }\textbf {\bibinfo {volume} {111}},\ \bibinfo {pages} {053524}
  (\bibinfo {year} {2012})}\BibitemShut {NoStop}%
\bibitem [{\citenamefont {Semenov}\ and\ \citenamefont
  {Kim}(2007)}]{Semenov2007}%
  \BibitemOpen
  \bibfield  {author} {\bibinfo {author} {\bibfnamefont {Y.~G.}\ \bibnamefont
  {Semenov}}\ and\ \bibinfo {author} {\bibfnamefont {K.~W.}\ \bibnamefont
  {Kim}},\ }\bibfield  {title} {\enquote {\bibinfo {title} {{Elastic
  spin-relaxation processes in semiconductor quantum dots}},}\ }\href {\doibase
  10.1103/PhysRevB.75.195342} {\bibfield  {journal} {\bibinfo  {journal} {Phys.
  Rev. B}\ }\textbf {\bibinfo {volume} {75}},\ \bibinfo {pages} {195342}
  (\bibinfo {year} {2007})}\BibitemShut {NoStop}%
\bibitem [{\citenamefont {Braun}\ \emph {et~al.}(2005)\citenamefont {Braun},
  \citenamefont {Marie}, \citenamefont {Lombez}, \citenamefont {Urbaszek},
  \citenamefont {Amand}, \citenamefont {Renucci}, \citenamefont {Kalevich},
  \citenamefont {Kavokin}, \citenamefont {Krebs}, \citenamefont {Voisin},\ and\
  \citenamefont {Masumoto}}]{Braun2005}%
  \BibitemOpen
  \bibfield  {author} {\bibinfo {author} {\bibfnamefont {P.-F.}\ \bibnamefont
  {Braun}}, \bibinfo {author} {\bibfnamefont {X.}~\bibnamefont {Marie}},
  \bibinfo {author} {\bibfnamefont {L.}~\bibnamefont {Lombez}}, \bibinfo
  {author} {\bibfnamefont {B.}~\bibnamefont {Urbaszek}}, \bibinfo {author}
  {\bibfnamefont {T.}~\bibnamefont {Amand}}, \bibinfo {author} {\bibfnamefont
  {P.}~\bibnamefont {Renucci}}, \bibinfo {author} {\bibfnamefont {V.~K.}\
  \bibnamefont {Kalevich}}, \bibinfo {author} {\bibfnamefont {K.~V.}\
  \bibnamefont {Kavokin}}, \bibinfo {author} {\bibfnamefont {O.}~\bibnamefont
  {Krebs}}, \bibinfo {author} {\bibfnamefont {P.}~\bibnamefont {Voisin}}, \
  and\ \bibinfo {author} {\bibfnamefont {Y.}~\bibnamefont {Masumoto}},\
  }\bibfield  {title} {\enquote {\bibinfo {title} {{Direct Observation of the
  Electron Spin Relaxation Induced by Nuclei in Quantum Dots}},}\ }\href
  {\doibase 10.1103/PhysRevLett.94.116601} {\bibfield  {journal} {\bibinfo
  {journal} {Phys. Rev. Lett.}\ }\textbf {\bibinfo {volume} {94}},\ \bibinfo
  {pages} {116601} (\bibinfo {year} {2005})}\BibitemShut {NoStop}%
\bibitem [{\citenamefont {Eble}\ \emph {et~al.}(2009)\citenamefont {Eble},
  \citenamefont {Testelin}, \citenamefont {Desfonds}, \citenamefont
  {Bernardot}, \citenamefont {Balocchi}, \citenamefont {Amand}, \citenamefont
  {Miard}, \citenamefont {Lema{\^{i}}tre}, \citenamefont {Marie},\ and\
  \citenamefont {Chamarro}}]{Eble2009}%
  \BibitemOpen
  \bibfield  {author} {\bibinfo {author} {\bibfnamefont {B.}~\bibnamefont
  {Eble}}, \bibinfo {author} {\bibfnamefont {C.}~\bibnamefont {Testelin}},
  \bibinfo {author} {\bibfnamefont {P.}~\bibnamefont {Desfonds}}, \bibinfo
  {author} {\bibfnamefont {F.}~\bibnamefont {Bernardot}}, \bibinfo {author}
  {\bibfnamefont {A.}~\bibnamefont {Balocchi}}, \bibinfo {author}
  {\bibfnamefont {T.}~\bibnamefont {Amand}}, \bibinfo {author} {\bibfnamefont
  {A.}~\bibnamefont {Miard}}, \bibinfo {author} {\bibfnamefont
  {A.}~\bibnamefont {Lema{\^{i}}tre}}, \bibinfo {author} {\bibfnamefont
  {X.}~\bibnamefont {Marie}}, \ and\ \bibinfo {author} {\bibfnamefont
  {M.}~\bibnamefont {Chamarro}},\ }\bibfield  {title} {\enquote {\bibinfo
  {title} {{Hole–Nuclear Spin Interaction in Quantum Dots}},}\ }\href
  {\doibase 10.1103/PhysRevLett.102.146601} {\bibfield  {journal} {\bibinfo
  {journal} {Phys. Rev. Lett.}\ }\textbf {\bibinfo {volume} {102}},\ \bibinfo
  {pages} {146601} (\bibinfo {year} {2009})}\BibitemShut {NoStop}%
\bibitem [{\citenamefont {Bechtold}\ \emph {et~al.}(2015)\citenamefont
  {Bechtold}, \citenamefont {Rauch}, \citenamefont {Li}, \citenamefont
  {Simmet}, \citenamefont {Ardelt}, \citenamefont {Regler}, \citenamefont
  {M{\"{u}}ller}, \citenamefont {Sinitsyn},\ and\ \citenamefont
  {Finley}}]{Bechtold2015}%
  \BibitemOpen
  \bibfield  {author} {\bibinfo {author} {\bibfnamefont {A.}~\bibnamefont
  {Bechtold}}, \bibinfo {author} {\bibfnamefont {D.}~\bibnamefont {Rauch}},
  \bibinfo {author} {\bibfnamefont {F.}~\bibnamefont {Li}}, \bibinfo {author}
  {\bibfnamefont {T.}~\bibnamefont {Simmet}}, \bibinfo {author} {\bibfnamefont
  {P.-L.}\ \bibnamefont {Ardelt}}, \bibinfo {author} {\bibfnamefont
  {A.}~\bibnamefont {Regler}}, \bibinfo {author} {\bibfnamefont {Kai}\
  \bibnamefont {M{\"{u}}ller}}, \bibinfo {author} {\bibfnamefont {N.~A.}\
  \bibnamefont {Sinitsyn}}, \ and\ \bibinfo {author} {\bibfnamefont {J.~J.}\
  \bibnamefont {Finley}},\ }\bibfield  {title} {\enquote {\bibinfo {title}
  {{Three-stage decoherence dynamics of an electron spin qubit in an optically
  active quantum dot}},}\ }\href {\doibase 10.1038/nphys3470} {\bibfield
  {journal} {\bibinfo  {journal} {Nat. Phys.}\ }\textbf {\bibinfo {volume}
  {11}},\ \bibinfo {pages} {1005} (\bibinfo {year} {2015})}\BibitemShut
  {NoStop}%
\bibitem [{\citenamefont {Belykh}\ \emph {et~al.}(2015)\citenamefont {Belykh},
  \citenamefont {Greilich}, \citenamefont {Yakovlev}, \citenamefont {Yacob},
  \citenamefont {Reithmaier}, \citenamefont {Benyoucef},\ and\ \citenamefont
  {Bayer}}]{Belykh2015}%
  \BibitemOpen
  \bibfield  {author} {\bibinfo {author} {\bibfnamefont {V.~V.}\ \bibnamefont
  {Belykh}}, \bibinfo {author} {\bibfnamefont {A.}~\bibnamefont {Greilich}},
  \bibinfo {author} {\bibfnamefont {D.~R.}\ \bibnamefont {Yakovlev}}, \bibinfo
  {author} {\bibfnamefont {M.}~\bibnamefont {Yacob}}, \bibinfo {author}
  {\bibfnamefont {J.~P.}\ \bibnamefont {Reithmaier}}, \bibinfo {author}
  {\bibfnamefont {M.}~\bibnamefont {Benyoucef}}, \ and\ \bibinfo {author}
  {\bibfnamefont {M.}~\bibnamefont {Bayer}},\ }\bibfield  {title} {\enquote
  {\bibinfo {title} {{Electron and hole g factors in InAs/InAlGaAs
  self-assembled quantum dots emitting at telecom wavelengths}},}\ }\href
  {\doibase 10.1103/PhysRevB.92.165307} {\bibfield  {journal} {\bibinfo
  {journal} {Phys. Rev. B}\ }\textbf {\bibinfo {volume} {92}},\ \bibinfo
  {pages} {165307} (\bibinfo {year} {2015})}\BibitemShut {NoStop}%
\bibitem [{\citenamefont {Belykh}\ \emph
  {et~al.}(2016{\natexlab{a}})\citenamefont {Belykh}, \citenamefont {Yakovlev},
  \citenamefont {Schindler}, \citenamefont {Zhukov}, \citenamefont {Semina},
  \citenamefont {Yacob}, \citenamefont {Reithmaier}, \citenamefont
  {Benyoucef},\ and\ \citenamefont {Bayer}}]{Belykh2016}%
  \BibitemOpen
  \bibfield  {author} {\bibinfo {author} {\bibfnamefont {V.~V.}\ \bibnamefont
  {Belykh}}, \bibinfo {author} {\bibfnamefont {D.~R.}\ \bibnamefont
  {Yakovlev}}, \bibinfo {author} {\bibfnamefont {J.~J.}\ \bibnamefont
  {Schindler}}, \bibinfo {author} {\bibfnamefont {E.~A.}\ \bibnamefont
  {Zhukov}}, \bibinfo {author} {\bibfnamefont {M.~A.}\ \bibnamefont {Semina}},
  \bibinfo {author} {\bibfnamefont {M.}~\bibnamefont {Yacob}}, \bibinfo
  {author} {\bibfnamefont {J.~P.}\ \bibnamefont {Reithmaier}}, \bibinfo
  {author} {\bibfnamefont {M.}~\bibnamefont {Benyoucef}}, \ and\ \bibinfo
  {author} {\bibfnamefont {M.}~\bibnamefont {Bayer}},\ }\bibfield  {title}
  {\enquote {\bibinfo {title} {{Large anisotropy of electron and hole g factors
  in infrared-emitting InAs/InAlGaAs self-assembled quantum dots}},}\ }\href
  {\doibase 10.1103/PhysRevB.93.125302} {\bibfield  {journal} {\bibinfo
  {journal} {Phys. Rev. B}\ }\textbf {\bibinfo {volume} {93}},\ \bibinfo
  {pages} {125302} (\bibinfo {year} {2016}{\natexlab{a}})}\BibitemShut
  {NoStop}%
\bibitem [{\citenamefont {van Bree}\ \emph {et~al.}(2016)\citenamefont {van
  Bree}, \citenamefont {Silov}, \citenamefont {van Maasakkers}, \citenamefont
  {Pryor}, \citenamefont {Flatt{\'{e}}},\ and\ \citenamefont
  {Koenraad}}]{VanBree2016}%
  \BibitemOpen
  \bibfield  {author} {\bibinfo {author} {\bibfnamefont {J.}~\bibnamefont {van
  Bree}}, \bibinfo {author} {\bibfnamefont {A.~Yu.}\ \bibnamefont {Silov}},
  \bibinfo {author} {\bibfnamefont {M.~L.}\ \bibnamefont {van Maasakkers}},
  \bibinfo {author} {\bibfnamefont {C.~E.}\ \bibnamefont {Pryor}}, \bibinfo
  {author} {\bibfnamefont {M.~E.}\ \bibnamefont {Flatt{\'{e}}}}, \ and\
  \bibinfo {author} {\bibfnamefont {P.~M.}\ \bibnamefont {Koenraad}},\
  }\bibfield  {title} {\enquote {\bibinfo {title} {{Anisotropy of electron and
  hole g tensors of quantum dots: An intuitive picture based on spin-correlated
  orbital currents}},}\ }\href {\doibase 10.1103/PhysRevB.93.035311} {\bibfield
   {journal} {\bibinfo  {journal} {Phys. Rev. B}\ }\textbf {\bibinfo {volume}
  {93}},\ \bibinfo {pages} {035311} (\bibinfo {year} {2016})}\BibitemShut
  {NoStop}%
\bibitem [{\citenamefont {Belykh}\ \emph
  {et~al.}(2016{\natexlab{b}})\citenamefont {Belykh}, \citenamefont {Yakovlev},
  \citenamefont {Schindler}, \citenamefont {van Bree}, \citenamefont
  {Koenraad}, \citenamefont {Averkiev}, \citenamefont {Bayer},\ and\
  \citenamefont {Silov}}]{Belykh2016a}%
  \BibitemOpen
  \bibfield  {author} {\bibinfo {author} {\bibfnamefont {V.~V.}\ \bibnamefont
  {Belykh}}, \bibinfo {author} {\bibfnamefont {D.~R.}\ \bibnamefont
  {Yakovlev}}, \bibinfo {author} {\bibfnamefont {J.~J.}\ \bibnamefont
  {Schindler}}, \bibinfo {author} {\bibfnamefont {J.}~\bibnamefont {van Bree}},
  \bibinfo {author} {\bibfnamefont {P.~M.}\ \bibnamefont {Koenraad}}, \bibinfo
  {author} {\bibfnamefont {N.~S.}\ \bibnamefont {Averkiev}}, \bibinfo {author}
  {\bibfnamefont {M.}~\bibnamefont {Bayer}}, \ and\ \bibinfo {author}
  {\bibfnamefont {A.~Yu.}\ \bibnamefont {Silov}},\ }\bibfield  {title}
  {\enquote {\bibinfo {title} {{Dispersion of the electron g factor anisotropy
  in InAs/InP self-assembled quantum dots}},}\ }\href {\doibase
  10.1063/1.4961201} {\bibfield  {journal} {\bibinfo  {journal} {J. Appl.
  Phys.}\ }\textbf {\bibinfo {volume} {120}},\ \bibinfo {pages} {084301}
  (\bibinfo {year} {2016}{\natexlab{b}})}\BibitemShut {NoStop}%
\bibitem [{\citenamefont {Syperek}\ \emph {et~al.}(2016)\citenamefont
  {Syperek}, \citenamefont {Dusanowski}, \citenamefont {Gawe{\l}czyk},
  \citenamefont {Sȩk}, \citenamefont {Somers}, \citenamefont {Reithmaier},
  \citenamefont {H{\"{o}}fling},\ and\ \citenamefont
  {Misiewicz}}]{Syperek2016}%
  \BibitemOpen
  \bibfield  {author} {\bibinfo {author} {\bibfnamefont {M.}~\bibnamefont
  {Syperek}}, \bibinfo {author} {\bibfnamefont {L.}~\bibnamefont {Dusanowski}},
  \bibinfo {author} {\bibfnamefont {M.}~\bibnamefont {Gawe{\l}czyk}}, \bibinfo
  {author} {\bibfnamefont {G.}~\bibnamefont {Sȩk}}, \bibinfo {author}
  {\bibfnamefont {A.}~\bibnamefont {Somers}}, \bibinfo {author} {\bibfnamefont
  {J.~P.}\ \bibnamefont {Reithmaier}}, \bibinfo {author} {\bibfnamefont
  {S.}~\bibnamefont {H{\"{o}}fling}}, \ and\ \bibinfo {author} {\bibfnamefont
  {J.}~\bibnamefont {Misiewicz}},\ }\bibfield  {title} {\enquote {\bibinfo
  {title} {{Exciton spin relaxation in InAs/InGaAlAs/InP(001) quantum dashes
  emitting near 1.55 $\mu$m}},}\ }\href {\doibase 10.1063/1.4966997} {\bibfield
   {journal} {\bibinfo  {journal} {Appl. Phys. Lett.}\ }\textbf {\bibinfo
  {volume} {109}},\ \bibinfo {pages} {193108} (\bibinfo {year}
  {2016})}\BibitemShut {NoStop}%
\bibitem [{\citenamefont {Merkulov}\ \emph {et~al.}(2002)\citenamefont
  {Merkulov}, \citenamefont {Efros},\ and\ \citenamefont
  {Rosen}}]{Merkulov2002}%
  \BibitemOpen
  \bibfield  {author} {\bibinfo {author} {\bibfnamefont {I.~A.}\ \bibnamefont
  {Merkulov}}, \bibinfo {author} {\bibfnamefont {Al.~L.}\ \bibnamefont
  {Efros}}, \ and\ \bibinfo {author} {\bibfnamefont {M.}~\bibnamefont
  {Rosen}},\ }\bibfield  {title} {\enquote {\bibinfo {title} {{Electron spin
  relaxation by nuclei in semiconductor quantum dots}},}\ }\href {\doibase
  10.1103/PhysRevB.65.205309} {\bibfield  {journal} {\bibinfo  {journal} {Phys.
  Rev. B}\ }\textbf {\bibinfo {volume} {65}},\ \bibinfo {pages} {205309}
  (\bibinfo {year} {2002})}\BibitemShut {NoStop}%
\bibitem [{\citenamefont {Smirnov}\ \emph {et~al.}(2018)\citenamefont
  {Smirnov}, \citenamefont {Zhukov}, \citenamefont {Kirstein}, \citenamefont
  {Yakovlev}, \citenamefont {Reuter}, \citenamefont {Wieck}, \citenamefont
  {Bayer}, \citenamefont {Greilich},\ and\ \citenamefont
  {Glazov}}]{Smirnov2018}%
  \BibitemOpen
  \bibfield  {author} {\bibinfo {author} {\bibfnamefont {D.~S.}\ \bibnamefont
  {Smirnov}}, \bibinfo {author} {\bibfnamefont {E.~A.}\ \bibnamefont {Zhukov}},
  \bibinfo {author} {\bibfnamefont {E.}~\bibnamefont {Kirstein}}, \bibinfo
  {author} {\bibfnamefont {D.~R.}\ \bibnamefont {Yakovlev}}, \bibinfo {author}
  {\bibfnamefont {D.}~\bibnamefont {Reuter}}, \bibinfo {author} {\bibfnamefont
  {A.~D.}\ \bibnamefont {Wieck}}, \bibinfo {author} {\bibfnamefont
  {M.}~\bibnamefont {Bayer}}, \bibinfo {author} {\bibfnamefont
  {A.}~\bibnamefont {Greilich}}, \ and\ \bibinfo {author} {\bibfnamefont
  {M.~M.}\ \bibnamefont {Glazov}},\ }\bibfield  {title} {\enquote {\bibinfo
  {title} {{Theory of spin inertia in singly charged quantum dots}},}\ }\href
  {\doibase 10.1103/PhysRevB.98.125306} {\bibfield  {journal} {\bibinfo
  {journal} {Phys. Rev. B}\ }\textbf {\bibinfo {volume} {98}},\ \bibinfo
  {pages} {125306} (\bibinfo {year} {2018})}\BibitemShut {NoStop}%
\bibitem [{\citenamefont {Yugova}\ \emph {et~al.}(2009)\citenamefont {Yugova},
  \citenamefont {Glazov}, \citenamefont {Ivchenko},\ and\ \citenamefont
  {Efros}}]{Yugova2009}%
  \BibitemOpen
  \bibfield  {author} {\bibinfo {author} {\bibfnamefont {I.~A.}\ \bibnamefont
  {Yugova}}, \bibinfo {author} {\bibfnamefont {M.~M.}\ \bibnamefont {Glazov}},
  \bibinfo {author} {\bibfnamefont {E.~L.}\ \bibnamefont {Ivchenko}}, \ and\
  \bibinfo {author} {\bibfnamefont {Al.~L.}\ \bibnamefont {Efros}},\ }\bibfield
   {title} {\enquote {\bibinfo {title} {{Pump-probe Faraday rotation and
  ellipticity in an ensemble of singly charged quantum dots}},}\ }\href
  {\doibase 10.1103/PhysRevB.80.104436} {\bibfield  {journal} {\bibinfo
  {journal} {Phys. Rev. B}\ }\textbf {\bibinfo {volume} {80}},\ \bibinfo
  {pages} {104436} (\bibinfo {year} {2009})}\BibitemShut {NoStop}%
\bibitem [{\citenamefont {Yugova}\ \emph {et~al.}(2007)\citenamefont {Yugova},
  \citenamefont {Greilich}, \citenamefont {Zhukov}, \citenamefont {Yakovlev},
  \citenamefont {Bayer}, \citenamefont {Reuter},\ and\ \citenamefont
  {Wieck}}]{Yugova2007}%
  \BibitemOpen
  \bibfield  {author} {\bibinfo {author} {\bibfnamefont {I.~A.}\ \bibnamefont
  {Yugova}}, \bibinfo {author} {\bibfnamefont {A.}~\bibnamefont {Greilich}},
  \bibinfo {author} {\bibfnamefont {E.~A.}\ \bibnamefont {Zhukov}}, \bibinfo
  {author} {\bibfnamefont {D.~R.}\ \bibnamefont {Yakovlev}}, \bibinfo {author}
  {\bibfnamefont {M.}~\bibnamefont {Bayer}}, \bibinfo {author} {\bibfnamefont
  {D.}~\bibnamefont {Reuter}}, \ and\ \bibinfo {author} {\bibfnamefont {A.~D.}\
  \bibnamefont {Wieck}},\ }\bibfield  {title} {\enquote {\bibinfo {title}
  {{Exciton fine structure in InGaAs∕GaAs quantum dots revisited by
  pump-probe Faraday rotation}},}\ }\href {\doibase 10.1103/PhysRevB.75.195325}
  {\bibfield  {journal} {\bibinfo  {journal} {Phys. Rev. B}\ }\textbf {\bibinfo
  {volume} {75}},\ \bibinfo {pages} {195325} (\bibinfo {year}
  {2007})}\BibitemShut {NoStop}%
\bibitem [{\citenamefont {Tartakovskii}\ \emph {et~al.}(2004)\citenamefont
  {Tartakovskii}, \citenamefont {Makhonin}, \citenamefont {Sellers},
  \citenamefont {Cahill}, \citenamefont {Andreev}, \citenamefont {Whittaker},
  \citenamefont {Wells}, \citenamefont {Fox}, \citenamefont {Mowbray},
  \citenamefont {Skolnick}, \citenamefont {Groom}, \citenamefont {Steer},
  \citenamefont {Liu},\ and\ \citenamefont {Hopkinson}}]{Tartakovskii2004}%
  \BibitemOpen
  \bibfield  {author} {\bibinfo {author} {\bibfnamefont {A.~I.}\ \bibnamefont
  {Tartakovskii}}, \bibinfo {author} {\bibfnamefont {M.~N.}\ \bibnamefont
  {Makhonin}}, \bibinfo {author} {\bibfnamefont {I.~R.}\ \bibnamefont
  {Sellers}}, \bibinfo {author} {\bibfnamefont {J.}~\bibnamefont {Cahill}},
  \bibinfo {author} {\bibfnamefont {A.~D.}\ \bibnamefont {Andreev}}, \bibinfo
  {author} {\bibfnamefont {D.~M.}\ \bibnamefont {Whittaker}}, \bibinfo {author}
  {\bibfnamefont {J-P.~R.}\ \bibnamefont {Wells}}, \bibinfo {author}
  {\bibfnamefont {A.~M.}\ \bibnamefont {Fox}}, \bibinfo {author} {\bibfnamefont
  {D.~J.}\ \bibnamefont {Mowbray}}, \bibinfo {author} {\bibfnamefont {M.~S.}\
  \bibnamefont {Skolnick}}, \bibinfo {author} {\bibfnamefont {K.~M.}\
  \bibnamefont {Groom}}, \bibinfo {author} {\bibfnamefont {M.~J.}\ \bibnamefont
  {Steer}}, \bibinfo {author} {\bibfnamefont {H.~Y.}\ \bibnamefont {Liu}}, \
  and\ \bibinfo {author} {\bibfnamefont {M.}~\bibnamefont {Hopkinson}},\
  }\bibfield  {title} {\enquote {\bibinfo {title} {{Effect of thermal annealing
  and strain engineering on the fine structure of quantum dot excitons}},}\
  }\href {\doibase 10.1103/PhysRevB.70.193303} {\bibfield  {journal} {\bibinfo
  {journal} {Phys. Rev. B}\ }\textbf {\bibinfo {volume} {70}},\ \bibinfo
  {pages} {193303} (\bibinfo {year} {2004})}\BibitemShut {NoStop}%
\bibitem [{\citenamefont {Belykh}\ \emph
  {et~al.}(2016{\natexlab{c}})\citenamefont {Belykh}, \citenamefont {Evers},
  \citenamefont {Yakovlev}, \citenamefont {Fobbe}, \citenamefont {Greilich},\
  and\ \citenamefont {Bayer}}]{Belykh2016b}%
  \BibitemOpen
  \bibfield  {author} {\bibinfo {author} {\bibfnamefont {V.~V.}\ \bibnamefont
  {Belykh}}, \bibinfo {author} {\bibfnamefont {E.}~\bibnamefont {Evers}},
  \bibinfo {author} {\bibfnamefont {D.~R.}\ \bibnamefont {Yakovlev}}, \bibinfo
  {author} {\bibfnamefont {F.}~\bibnamefont {Fobbe}}, \bibinfo {author}
  {\bibfnamefont {A.}~\bibnamefont {Greilich}}, \ and\ \bibinfo {author}
  {\bibfnamefont {M.}~\bibnamefont {Bayer}},\ }\bibfield  {title} {\enquote
  {\bibinfo {title} {{Extended pump-probe Faraday rotation spectroscopy of the
  submicrosecond electron spin dynamics in n-type GaAs}},}\ }\href {\doibase
  10.1103/PhysRevB.94.241202} {\bibfield  {journal} {\bibinfo  {journal} {Phys.
  Rev. B}\ }\textbf {\bibinfo {volume} {94}},\ \bibinfo {pages} {241202(R)}
  (\bibinfo {year} {2016}{\natexlab{c}})}\BibitemShut {NoStop}%
\bibitem [{\citenamefont {Zhukov}\ \emph {et~al.}(2018)\citenamefont {Zhukov},
  \citenamefont {Kirstein}, \citenamefont {Smirnov}, \citenamefont {Yakovlev},
  \citenamefont {Glazov}, \citenamefont {Reuter}, \citenamefont {Wieck},
  \citenamefont {Bayer},\ and\ \citenamefont {Greilich}}]{Zhukov2018}%
  \BibitemOpen
  \bibfield  {author} {\bibinfo {author} {\bibfnamefont {E.~A.}\ \bibnamefont
  {Zhukov}}, \bibinfo {author} {\bibfnamefont {E.}~\bibnamefont {Kirstein}},
  \bibinfo {author} {\bibfnamefont {D.~S.}\ \bibnamefont {Smirnov}}, \bibinfo
  {author} {\bibfnamefont {D.~R.}\ \bibnamefont {Yakovlev}}, \bibinfo {author}
  {\bibfnamefont {M.~M.}\ \bibnamefont {Glazov}}, \bibinfo {author}
  {\bibfnamefont {D.}~\bibnamefont {Reuter}}, \bibinfo {author} {\bibfnamefont
  {A.~D.}\ \bibnamefont {Wieck}}, \bibinfo {author} {\bibfnamefont
  {M.}~\bibnamefont {Bayer}}, \ and\ \bibinfo {author} {\bibfnamefont
  {A.}~\bibnamefont {Greilich}},\ }\bibfield  {title} {\enquote {\bibinfo
  {title} {{Spin inertia of resident and photoexcited carriers in singly
  charged quantum dots}},}\ }\href {\doibase 10.1103/PhysRevB.98.121304}
  {\bibfield  {journal} {\bibinfo  {journal} {Phys. Rev. B}\ }\textbf {\bibinfo
  {volume} {98}},\ \bibinfo {pages} {121304} (\bibinfo {year}
  {2018})}\BibitemShut {NoStop}%
\bibitem [{\citenamefont {Kikkawa}\ and\ \citenamefont
  {Awschalom}(1998)}]{Kikkawa1998}%
  \BibitemOpen
  \bibfield  {author} {\bibinfo {author} {\bibfnamefont {J.~M.}\ \bibnamefont
  {Kikkawa}}\ and\ \bibinfo {author} {\bibfnamefont {D.~D.}\ \bibnamefont
  {Awschalom}},\ }\bibfield  {title} {\enquote {\bibinfo {title} {{Resonant
  Spin Amplification in n-Type GaAs}},}\ }\href {\doibase
  10.1103/PhysRevLett.80.4313} {\bibfield  {journal} {\bibinfo  {journal}
  {Phys. Rev. Lett.}\ }\textbf {\bibinfo {volume} {80}},\ \bibinfo {pages}
  {4313} (\bibinfo {year} {1998})}\BibitemShut {NoStop}%
\bibitem [{\citenamefont {Heisterkamp}\ \emph {et~al.}(2015)\citenamefont
  {Heisterkamp}, \citenamefont {Zhukov}, \citenamefont {Greilich},
  \citenamefont {Yakovlev}, \citenamefont {Korenev}, \citenamefont {Pawlis},\
  and\ \citenamefont {Bayer}}]{Heisterkamp2015}%
  \BibitemOpen
  \bibfield  {author} {\bibinfo {author} {\bibfnamefont {F.}~\bibnamefont
  {Heisterkamp}}, \bibinfo {author} {\bibfnamefont {E.~A.}\ \bibnamefont
  {Zhukov}}, \bibinfo {author} {\bibfnamefont {A.}~\bibnamefont {Greilich}},
  \bibinfo {author} {\bibfnamefont {D.~R.}\ \bibnamefont {Yakovlev}}, \bibinfo
  {author} {\bibfnamefont {V.~L.}\ \bibnamefont {Korenev}}, \bibinfo {author}
  {\bibfnamefont {A.}~\bibnamefont {Pawlis}}, \ and\ \bibinfo {author}
  {\bibfnamefont {M.}~\bibnamefont {Bayer}},\ }\bibfield  {title} {\enquote
  {\bibinfo {title} {{Longitudinal and transverse spin dynamics of donor-bound
  electrons in fluorine-doped ZnSe: Spin inertia versus Hanle effect}},}\
  }\href {\doibase 10.1103/PhysRevB.91.235432} {\bibfield  {journal} {\bibinfo
  {journal} {Phys. Rev. B}\ }\textbf {\bibinfo {volume} {91}},\ \bibinfo
  {pages} {235432} (\bibinfo {year} {2015})}\BibitemShut {NoStop}%
\bibitem [{\citenamefont {Golovach}\ \emph {et~al.}(2004)\citenamefont
  {Golovach}, \citenamefont {Khaetskii},\ and\ \citenamefont
  {Loss}}]{Golovach2004}%
  \BibitemOpen
  \bibfield  {author} {\bibinfo {author} {\bibfnamefont {V.~N.}\ \bibnamefont
  {Golovach}}, \bibinfo {author} {\bibfnamefont {A.}~\bibnamefont {Khaetskii}},
  \ and\ \bibinfo {author} {\bibfnamefont {D.}~\bibnamefont {Loss}},\
  }\bibfield  {title} {\enquote {\bibinfo {title} {{Phonon-Induced Decay of the
  Electron Spin in Quantum Dots}},}\ }\href {\doibase
  10.1103/PhysRevLett.93.016601} {\bibfield  {journal} {\bibinfo  {journal}
  {Phys. Rev. Lett.}\ }\textbf {\bibinfo {volume} {93}},\ \bibinfo {pages}
  {016601} (\bibinfo {year} {2004})}\BibitemShut {NoStop}%
\bibitem [{\citenamefont {Tsitsishvili}\ \emph {et~al.}(2002)\citenamefont
  {Tsitsishvili}, \citenamefont {Baltz},\ and\ \citenamefont
  {Kalt}}]{Tsitsishvili2002Temp}%
  \BibitemOpen
  \bibfield  {author} {\bibinfo {author} {\bibfnamefont {E.}~\bibnamefont
  {Tsitsishvili}}, \bibinfo {author} {\bibfnamefont {R.~V.}\ \bibnamefont
  {Baltz}}, \ and\ \bibinfo {author} {\bibfnamefont {H.}~\bibnamefont {Kalt}},\
  }\bibfield  {title} {\enquote {\bibinfo {title} {{Temperature dependence of
  polarization relaxation in semiconductor quantum dots}},}\ }\href {\doibase
  10.1103/PhysRevB.66.161405} {\bibfield  {journal} {\bibinfo  {journal} {Phys.
  Rev. B}\ }\textbf {\bibinfo {volume} {66}},\ \bibinfo {pages} {161405}
  (\bibinfo {year} {2002})}\BibitemShut {NoStop}%
\bibitem [{\citenamefont {Flissikowski}\ \emph {et~al.}(2003)\citenamefont
  {Flissikowski}, \citenamefont {Akimov}, \citenamefont {Hundt},\ and\
  \citenamefont {Henneberger}}]{Flissikowski2003}%
  \BibitemOpen
  \bibfield  {author} {\bibinfo {author} {\bibfnamefont {T.}~\bibnamefont
  {Flissikowski}}, \bibinfo {author} {\bibfnamefont {I.~A.}\ \bibnamefont
  {Akimov}}, \bibinfo {author} {\bibfnamefont {A.}~\bibnamefont {Hundt}}, \
  and\ \bibinfo {author} {\bibfnamefont {F.}~\bibnamefont {Henneberger}},\
  }\bibfield  {title} {\enquote {\bibinfo {title} {{Single-hole spin relaxation
  in a quantum dot}},}\ }\href {\doibase 10.1103/PhysRevB.68.161309} {\bibfield
   {journal} {\bibinfo  {journal} {Phys. Rev. B}\ }\textbf {\bibinfo {volume}
  {68}},\ \bibinfo {pages} {161309} (\bibinfo {year} {2003})}\BibitemShut
  {NoStop}%
\bibitem [{\citenamefont {Khaetskii}\ and\ \citenamefont
  {Nazarov}(2001)}]{Khaetskii2001}%
  \BibitemOpen
  \bibfield  {author} {\bibinfo {author} {\bibfnamefont {A.~V.}\ \bibnamefont
  {Khaetskii}}\ and\ \bibinfo {author} {\bibfnamefont {Y.~V.}\ \bibnamefont
  {Nazarov}},\ }\bibfield  {title} {\enquote {\bibinfo {title} {{Spin-flip
  transitions between Zeeman sublevels in semiconductor quantum dots}},}\
  }\href {\doibase 10.1103/PhysRevB.64.125316} {\bibfield  {journal} {\bibinfo
  {journal} {Phys. Rev. B}\ }\textbf {\bibinfo {volume} {64}},\ \bibinfo
  {pages} {125316} (\bibinfo {year} {2001})}\BibitemShut {NoStop}%
\bibitem [{\citenamefont {Erlingsson}\ \emph {et~al.}(2001)\citenamefont
  {Erlingsson}, \citenamefont {Nazarov},\ and\ \citenamefont
  {Fal'ko}}]{Erlingsson2001}%
  \BibitemOpen
  \bibfield  {author} {\bibinfo {author} {\bibfnamefont {S.~I.}\ \bibnamefont
  {Erlingsson}}, \bibinfo {author} {\bibfnamefont {Y.~V.}\ \bibnamefont
  {Nazarov}}, \ and\ \bibinfo {author} {\bibfnamefont {V.~I.}\ \bibnamefont
  {Fal'ko}},\ }\bibfield  {title} {\enquote {\bibinfo {title}
  {{Nucleus-mediated spin-flip transitions in GaAs quantum dots}},}\ }\href
  {\doibase 10.1103/PhysRevB.64.195306} {\bibfield  {journal} {\bibinfo
  {journal} {Phys. Rev. B}\ }\textbf {\bibinfo {volume} {64}},\ \bibinfo
  {pages} {195306} (\bibinfo {year} {2001})}\BibitemShut {NoStop}%
\bibitem [{\citenamefont {Abalmassov}\ and\ \citenamefont
  {Marquardt}(2004)}]{Abalmassov2004}%
  \BibitemOpen
  \bibfield  {author} {\bibinfo {author} {\bibfnamefont {V.~A.}\ \bibnamefont
  {Abalmassov}}\ and\ \bibinfo {author} {\bibfnamefont {F.}~\bibnamefont
  {Marquardt}},\ }\bibfield  {title} {\enquote {\bibinfo {title}
  {{Electron-nuclei spin relaxation through phonon-assisted hyperfine
  interaction in a quantum dot}},}\ }\href {\doibase
  10.1103/PhysRevB.70.075313} {\bibfield  {journal} {\bibinfo  {journal} {Phys.
  Rev. B}\ }\textbf {\bibinfo {volume} {70}},\ \bibinfo {pages} {075313}
  (\bibinfo {year} {2004})}\BibitemShut {NoStop}%
\end{thebibliography}
\end{document}